\documentclass[twocolumn,superscriptaddress,showpacs,pra,floatfix]{revtex4}
\usepackage{graphicx}
\usepackage{amsmath}
\usepackage{amssymb}
\usepackage{amsfonts}

\usepackage{bm}

\newcommand{\mc}{\mathcal} 

\newcommand{\eqname}[1]{\label{eq:#1}}
\newcommand{\bgar}{\begin{eqnarray}}
\newcommand{\enar}[1]{\label{eq:#1}\end{eqnarray}}

\newcommand{\kk}{ {\bf k}}

\newcommand{\xx}{ {\bf x}}

\newcommand{\eq}[1]{(\ref{eq:#1})}

\newcommand{\psihd}{\hat\psi^\dagger}

\newcommand{\psih}{\hat\psi}

\newcommand{\RR}{{\bf R}}

\newcommand{\ahd}{\hat a^\dagger}
\newcommand{\ah}{\hat a}

\newcommand{\phd}{\hat p^\dagger}
\newcommand{\ph}{\hat p}

\newcommand{\alphah}{ {\hat \alpha}}

\begin{document}

\title{Optical properties of atomic Mott insulators: from slow light to dynamical Casimir effects}

%\affiliation{BEC-INFM, Universit\`a di Trento, 38050 Povo, Italy}
%\affiliation{Laboratoire Kastler Brossel, \'Ecole Normale
%Sup\'erieure, 24 rue Lhomond, 75005 Paris, France}

\author{Iacopo Carusotto}
\email{carusott@science.unitn.it}

\affiliation{Dipartimento di Fisica,  Universit\`a di Trento and CNR-INFM BEC Center, via Sommarive 14, I-38050 Povo, Italy} 

\author{Mauro Antezza}
\affiliation{Dipartimento di Fisica,  Universit\`a di Trento and CNR-INFM BEC Center, via Sommarive 14, I-38050 Povo, Italy} 

\author{Francesco Bariani}
\affiliation{Dipartimento di Fisica,  Universit\`a di Trento and CNR-INFM BEC Center, via Sommarive 14, I-38050 Povo, Italy} 

\author{Simone De Liberato}
\affiliation{Laboratoire Mat\'eriaux et Ph\'enom\`enes Quantiques, Universit\'e Paris Diderot-Paris 7 and CNRS, UMR7162, \\  B\^atiment Condorcet, 10 rue A. Domon et L. Duquet, 75013 Paris, France} 
\affiliation{Laboratoire Pierre Aigrain,
\'Ecole Normale Sup\'erieure, 24 rue Lhomond, 75005 Paris, France}

\author{Cristiano Ciuti}
\affiliation{Laboratoire Mat\'eriaux et Ph\'enom\`enes Quantiques, Universit\'e Paris Diderot-Paris 7 and CNRS, UMR7162, \\  B\^atiment Condorcet, 10 rue A. Domon et L. Duquet, 75013 Paris, France}

\begin{abstract}
We theoretically study the optical properties of a gas of ultracold, coherently dressed three-level atoms in a Mott insulator phase of an optical lattice.
The vacuum state, the band dispersion and the absorption spectrum of the polariton field can be controlled in real time by varying the amplitude and the frequency of the dressing beam.
In the weak dressing regime, the system shows unique ultra-slow light propagation properties without absorption.
In the presence of a fast time modulation of the dressing amplitude, we predict a significant emission of photon pairs by parametric amplification of the polaritonic zero-point fluctuations.
Quantitative considerations on the experimental observability of such a dynamical Casimir effect are presented for the most promising atomic species and level schemes.
\end{abstract}

%\vspace{1cm}

\pacs{
42.50.Nn, % Quantum optical phenomena in absorbing, amplifying, dispersive and conducting media; cooperative phenomena in quantum optical systems \\
03.75.Lm, % Tunneling, Josephson effect, Bose-Einstein condensates in periodic potentials, solitons, vortices, and topological excitations \\
71.36.+c, % Polaritons \\
03.70.+k %Theory of quantized fields
}

\date{\today}

\maketitle

\section{Introduction}

Most of the recent advances in the field of nonlinear and quantum optics were made possible by the development of novel optical media with unprecedented properties.
On one hand, the optical response of carriers in solid-state materials can be controlled and enhanced by confining the carrier motion and/or the photon mode in suitably grown nanostructures~\cite{Burstein,Microcav,CQED,ISB-QW-exp,cavity_enh}.
On the other hand, systems of ultracold atoms appear as very promising in view of all those applications which require long coherence times, e.g. quantum information processing.

Even though the low density of an atomic gas limits the absolute strength of the light-matter coupling, still these systems have the advantage of being almost immune from disorder and decoupled from the environment.
Furthermore, they offer the possibility of a precise control and wide tunability of the system parameters in real time by optical and/or magnetic means.
In particular, Mott~\cite{MI_exp,MI_th} (as well as band~\cite{band_I}) insulator states have been realized, where a constant and integer number of atoms are trapped in the extremely regular potential of an optical lattice.
Such systems constitute an almost perfect realization of the Hopfield model of resonant dielectrics~\cite{Hopfield}.

In the present paper we present a theoretical study of the classical and quantum optical properties of an atomic Mott insulator. Recently, the case of two-level atoms was investigated in~\cite{MI_optics,Zoubi}. Here we extend the Hopfield model to the case of a three-level system in the presence of a coherent dressing field.
The rich potential of three-level configurations has already been demonstrated with the observation of a variety of remarkable effects, such as the quenching of resonant absorption by the so-called  electromagnetically induced transparency (EIT) effect~\cite{Arimondo,EIT_fleisch}, the light propagation at ultra-slow group velocities in the m/s range~\cite{SlowLightReview,SlowLightExp}, and the coherent stopping and storing of light pulses~\cite{stop}.
Here we show how the peculiarities of atomic Mott insulator states can lead to further improvements of these experiments and, even more remarkably, open the way to studies of more subtle quantum optical effects.

Even in its ground state, the electromagnetic (e.m) field possesses in fact zero-point fluctuations, whose properties are non-trivially affected by the presence of dielectric and/or metallic bodies. One of the most celebrated consequence is the (static) Casimir effect, i.e. the appearance of a force between macroscopic objects due to the zero-point energy of the electromagnetic field~\cite{static,static_exp}. In the last decades, this force has been the object of intense experimental and theoretical studies in a number of different systems and its main properties can nowadays be considered as reasonably well understood.

The situation is completely different for what concerns the so-called {\em dynamical Casimir effect} (DCE)~\cite{DCE,Kardar}, i.e. the observable radiation that is emitted by the parametric excitation of the quantum vacuum when the boundary conditions and/or the propagation constants of the electromagnetic field are modulated in time on a very fast time scale.
In spite of a wide theoretical literature having addressed this effect for a variety of systems and excitation schemes~\cite{lambrecht,DCE_n,DCE_cond,PD,dezael,DCE_atomi}, no experimental observation has been reported yet, mainly because of the difficulty of modulating the system parameters at a high enough speed and the presence of competing spurious effects.

In the second part of this paper we show how Mott insulators of coherently dressed three-level atoms are very promising candidates for an experimental observation of the dynamical Casimir effect.
As the atomic response to e.m. fields depends strongly on the amplitude and the frequency of the dressing field, a significant time modulation of the optical properties of the atomic Mott insulator can be induced by modulating the dressing parameters on a very fast time scale via standard pulse manipulation techniques~\cite{fast-modul}.
On the other hand, the cleanness of atomic Mott insulator systems allows one to squeeze linewidths down to the spontaneous radiative level and hence to avoid those inhomogeneous broadening mechanisms that have so far limited the performances of slow- and stopped-light experiments.
Moreover, as the dynamical Casimir radiation is collected in the optical domain and no relaxation processes are involved in the modulation process, no difficulties are expected to appear as a consequence of thermal black-body radiation or incoherent luminescence from photoexcited carriers~\cite{PD}.

An {\em ab initio} model is developed to confirm these expectations in a quantitative way: taking inspiration from recent works~\cite{Saito,ISB-QW-th,ISB-QW-th_PRL}, we build a microscopic theory of the dynamical Casimir effect in atomic Mott insulators which explicitely includes the matter degrees of freedom. Thanks to the simple form of the resulting parametric Hamiltonian and to the relative weakness of the light-matter coupling constant, expressions for the emission intensity are obtained in closed analytical form. 
As expected, the most favorable frequency region appears to be the middle polariton branch (the so-called dark-state polariton of~\cite{fleischhauer}), which shows a strong resonant coupling of light with the matter degrees of freedom, as well as a still acceptable amount of absorption losses.
Advantages and disadvantages of the atomic Mott insulator system over previously studied solid state systems~\cite{ISB-QW-th,ISB-QW-th_PRL} will then be pointed out, as well as the criteria for the choice of the atomic levels to be used.
Quantitative estimations of the dynamical Casimir intensity in realistic systems appear as very promising in view of the experimental observation of this still elusive effect.
Generalization of the results to experimentally less demanding atomic states, e.g. Bose-condensed clouds or thermal gases is finally discussed.

The structure of the paper is the following. In Sec.\ref{sec:model} we introduce the physical system and the model used for its theoretical description. In Sec.\ref{sec:stationary} we discuss in a systematic way the static properties of the system, such as the dispersion and lifetime of the elementary excitations of the system, the so-called {\em polaritons}. A calculation of the dynamical Casimir emission in a spatially infinite, bulk system is presented in Sec.\ref{sec:time-dep}, and then extended to experimentally more relevant finite-size geometries in Sec.\ref{sec:realistic}. A quantitative discussion of the emission is presented in Sec.\ref{sec:quant} using realistic parameters of state-of-the-art samples. Conclusions are finally drawn in Sec.\ref{sec:conclu}.

\section{The physical system and the theoretical model}
\label{sec:model}

We consider a gas of atoms trapped in the periodic potential of a three-dimensional optical lattice with simple cubic geometry of lattice spacing $a_L$. Unless otherwise specified, the system is assumed to be spatially homogeneous with periodic boundary conditions in all three dimensions. The box sizes are equal to $L_{x,y,z}$, and the total volume of the system is $\mc{V}=L_xL_yL_z$.
For a tight enough lattice potential and commensurate filling, the ground state of the system corresponds to an Mott insulator state, with an integer number $n$ of atoms at each lattice site~\cite{MI_exp,MI_th} and almost negligible number fluctuations.
In what follows we focus our attention on the $n=1$ case in which the atoms are spatially separated and do not interact but via the electromagnetic field. The total number of atoms in the system is thus $N=L_xL_yL_z/a_L^3$ and the average density $n_{at}=1/a_L^3$. The temperature of the system is assumed to be low enough for the zero temperature approximation to hold.

\begin{figure}[htbp]
\begin{center}
\includegraphics[width=0.9\columnwidth,clip]{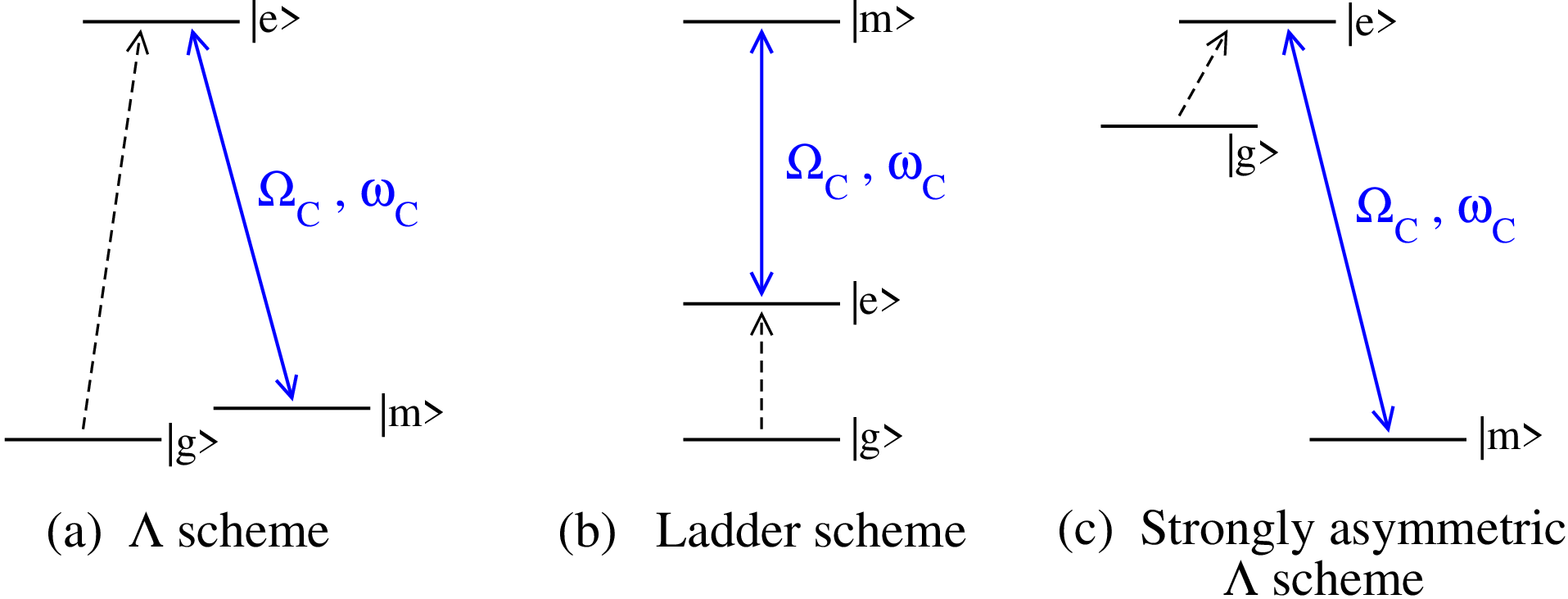}
\caption{Sketch of the level schemes under consideration.}
\label{fig:levels}
\end{center}
\end{figure}

The internal atomic dynamics takes place among three internal levels organized in either a $\Lambda$ or a ladder structure~\cite{Arimondo} as sketched in Fig.\ref{fig:levels}(a,b) (the case of a strongly asymmetric $\Lambda$ scheme of Fig.\ref{fig:levels}(c) will be discussed in Sec.\ref{sec:quant}). The atoms are initially prepared in their internal ground state $g$, which is connected to an excited state $e$ by an allowed optical transition of dipole matrix element $d_{eg}$ at a (bare) frequency $\omega_e$. For notational simplicity, the energy zero is set in a way to have $\omega_g=0$.
A coupling laser of frequency $\omega_C$ dresses the atoms by driving the transition between the excited $e$ level and a third, initially empty, state $m$ of energy $\omega_m$. In the $\Lambda$ case, the lifetime of the $m$ state can be very long, much longer than the free-space radiative lifetime of the $e$ state.

In terms of the local amplitude $E_C(\RR)$ of the dressing electric field at the atomic position $\RR$, the (complex) Rabi frequency of the coupling is $\Omega_C(\RR)=d_{em}\,E_C(\RR)$. 
In the following we consider the case where $\Omega_C$ is spatially uniform~\cite{larocca_sinus}; the discussion of more complex cases is postponed to future works.
The direct transition $g\rightarrow m$ is assumed to be optically inactive $d_{gm}\simeq 0$. 
Provided the lattice potential is strong enough to fulfill the Lamb-Dicke condition and has the same effect on the atoms irrespectively of their internal state, the external degrees of freedom can be decoupled from the internal dynamics and remain frozen in the motional ground state of each lattice site~\cite{cooling_review,CCT4}.

\subsection{The three-level Hopfield model}

A quantitative description of the many-atom system interacting with the electromagnetic field can be developed by generalizing the Hopfield model of a resonant dielectric~\cite{Hopfield} to the present case of three-level atoms. In this approach, both the atomic electric dipole polarization and the radiation field are described as a collection of coupled harmonic oscillators.
Neglecting for simplicity the photon polarization as well as all higher-lying photonic bands, the vector potential operator has a simple ``scalar'' expression in terms of the photon ($ph$) creation and annihilation operators $\ahd_{ph,\kk}$ and $\ah_{ph,\kk}$:
\begin{equation}
\hat{A}(\RR)=\sum_{\kk\in fBz} \sqrt{\frac{2\pi c\hbar}{k\,\mc{V}}}\, \left(\ah_{ph,\kk} e^{i\kk\RR} +
\ahd_{ph,\kk}  e^{-i\kk\RR} \right),
\end{equation}
where the sum over $\kk$ vectors is limited to the first Brillouin zone (fBz) of the lattice.
This approximation holds under the assumption of an isotropic atomic response and in the limit of a small lattice spacing $\omega_e a_L /c \ll 1$~\cite{bariani}.

In our specific case of $3$-level atoms, two material degrees of freedom are associated to every atom $j$, which correspond to its excitation from the $g$ state to respectively the $e$ and $m$ states. As usual, raising and lowering operators are defined as $\ahd_{e,j}\,|g\rangle_j=|e\rangle_j$ and $\ah_{e,j}\,|e\rangle_j=|g\rangle_j$ for the excited $e$ state, and analogously the $\ahd_{m,j}$ and $\ah_{m,j}$ for the $m$ state.
In the spirit of the harmonic oscillator model of \cite{Hopfield}, these operators can be extended as creation and annihilation operators satisfying the usual Bose commutation rules.
This bosonic description is accurate under the assumption that the probability for a given atom to be in an excited state is small
\footnote{The commutator of raising $a^\dagger_{e,j} = |e\rangle_j \langle g|_j$ and lowering $a_{e,j}=|g\rangle_j \langle e|_j$ operators is 
$[a_{e,j}, a^\dagger_{e,j}] = |g\rangle_j \langle g|_j-|e\rangle_j \langle e|_j=
\mathbf{1}_j-|m\rangle_j \langle m|_j-2\,|e\rangle_j \langle e|_j$, where $\mathbf{1}_j$ is the identity operator on the $j$th atom.
If the occupation of both $e$ and $m$ states is much smaller than 1, this commutator is well approximated by the identity $\mathbf{1}_j$ and thus has bosonic nature.}:
in this limit the higher-lying atomic states are not involved in the physics under consideration and the dynamics is taking place mostly in the subspace spanned by the three $|(g,e,m)\rangle$ states for which the harmonic oscillator description is exact. Generally speaking, this assumption is expected to be accurate as long as the number of excitations present in the system is much smaller than the number of atoms~\cite{Bose}.

Reabsorbing the phase of the dressing field $\Omega_C$ and its time-dependence at $\omega_C$ into the definition of the $\ah_{m,j}$ and $\ahd_{m,j}$ operators, the internal dynamics of the atom is described by the following time-independent Hamiltonian:
\begin{multline}
H_{at}^{j}=\hbar \omega_e \ahd_{e,j}\ah_{e,j}+
\hbar \tilde{\omega}_m\,\ahd_{m,j}\ah_{m,j}+\\
+\hbar \Omega_C\,\left(\ahd_{e,j}\ah_{m,j}+\ahd_{m,j}\ah_{e,j} \right)
\eqname{H_at_0}
\end{multline}
with a real $\Omega_C$ and a renormalized $\tilde{\omega}_m=\omega_m\pm \omega_C$, the $\pm$ signs referring to respectively the $\Lambda$ and the ladder configuration (see Fig.\ref{fig:levels}).

The electric-dipole coupling of the $g\rightarrow e$ transition to the transverse e.m. field
\footnote{Coupling to the longitudinal e.m. field corresponds to the static dipole-dipole Coulomb interaction~\cite{Hopfield,CCT4}.
At low excitation regimes, this term is only responsible for a slight red-shift of the $g\rightarrow e$ transition frequency $\omega_e$ by $\Delta\omega_e=-2\,\bar{C}^2/(3\omega_e)$. For the typical values of the systems under examination here, this shift is on the order of a few $10\,\textrm{MHz}$ and can be reincorporated in the definition of $\omega_e$.
In a semi-classical description of light-matter interaction, such a shift naturally appears when the Clausius-Mossotti form of the dielectric constant is used~\cite{Jackson,Agarwal}.
}

is included by means of the standard minimal coupling replacement.
The Hamiltonian \eq{H_at_0} can be in fact rewritten in terms of the harmonic oscillator position and momentum operators as
\begin{multline}
H_{at}^{j}=
\frac{M\,\omega_e^2}{2}\,\hat{X}_{e,j}^2+
\frac{1}{2M}\,\hat{P}_{e,j}^2+ \\
+\frac{M\,\tilde{\omega}_m^2}{2}\,\hat{X}_{m,j}^2+
\frac{1}{2M}\,\hat{P}_{m,j}^2+ \\
+M\Omega_C\sqrt{\omega_e\tilde{\omega}_m}\,\hat{X}_{e,j}\,\hat{X}_{m,j}+
\frac{\Omega_C}{M\sqrt{\omega_e\tilde{\omega}_m}}\,\hat{P}_{e,j}\hat{P}_{m,j}
\eqname{H_XP} 
\end{multline}
where
\begin{eqnarray}
{\hat X}_{e,j}&=&\sqrt{\frac{\hbar}{2M\omega_e}}\,\left(\ah_{e,j}+\ahd_{e,j}\right) \\
{\hat P}_{e,j}&=&i\sqrt{\frac{\hbar M\omega_e}{2}}\,\left(\ahd_{e,j}-\ah_{e,j}\right),
%{\hat %X}_{m,j}&=&\sqrt{\frac{\hbar}{2M\tilde{\omega}_m}}\,\left(\ah_{m,j}+\ahd_{m,j}\right)
%{\hat P}_{m,j}&=&i\sqrt{\frac{\hbar %M\tilde{\omega}_m}{2}}\,\left(\ahd_{m,j}-\ah_{m,j}\right).
\end{eqnarray}
Analogous expressions hold for ${\hat X}_{m,j}$ and ${\hat P}_{m,j}$, where $\ah_{e,j}$ is replaced by $\ah_{m,j}$ and $\omega_e$ by $\tilde{\omega}_m$.
The minimal coupling Hamiltonian is then obtained by replacing $\hat{P}_{e,j}$ with $\hat{P}_{e,j}-q\, \hat{A}(\RR_j)/c$ in \eq{H_XP}, while $\hat{P}_{m,j}$ is not affected because the $g\rightarrow m$ transition is optically forbidden. $\RR_j$ is here the position of the $j$ atom.

Note that the ``charge" $q$ and ``mass" $M$ parameters that appear in the harmonic oscillator model do not have a physical meaning {\em per se}, but are model parameters that are to be chosen in a way to reproduce the physics of the system under examination.
By comparing the dipole moment of the atomic transition to the matrix element between the ground and the first excited state of the harmonic oscillator model, one finds that they have to satisfy the relation $d_{eg}=\sqrt{\hbar q^2/2M\omega_e}$. All observable physical quantities will in fact involve only $d_{eg}$ and not the $q$ and $M$ separately.

To take full advantage of the translational symmetry of the system, it is useful to introduce the collective atomic operators
\begin{equation}
\ahd_{(e,m),\kk}=\frac{1}{\sqrt{N}} \sum_j \ahd_{(e,m),j}\,e^{i\kk\RR_j}
\end{equation}
which create a delocalized atomic excitation with a wavevector $\kk$ belonging to the first Brillouin zone of the lattice. Analogously to their localized counterparts $\ah_{(e,m),j}$ and $\ahd_{(e,m),j}$, the $\ah_{(e,m),\kk}$ and $\ahd_{(e,m),\kk}$ satisfy Bose commutation rules.

Straightforward manipulations lead to the final form of the light-matter Hamiltonian:
\begin{equation}
H=\sum_\kk \left[H_{ph,\kk}+H_{at,\kk}+H_{int,\kk}\right],
\eqname{H_tot}
\end{equation}
where
\begin{eqnarray}
H_{ph,\kk}&\!=\!&\hbar c k \left(\ahd_{ph,\kk} \ah_{ph,\kk} +
  \frac{1}{2}\right), \eqname{H_field} \\
H_{at,\kk}&\!=\!&\hbar \omega_e \ahd_{e,\kk}\ah_{e,\kk}+ 
\tilde{\omega}_m \ahd_{m,\kk}\ah_{m,\kk} \nonumber \\
&\!+\!& \hbar \Omega_C \,\left(\ahd_{e,\kk}\ah_{m,\kk} + \ahd_{m,\kk}
\ah_{e,\kk}\right), \eqname{H_at} \\
H_{int,\kk}&\!=\!&-i \hbar C_k \left(\ahd_{e,-\kk} -\ah_{e,\kk}\right)\,\left(\ah_{ph,-\kk}+\ahd_{ph,\kk}\right) \nonumber \\ &+&\hbar D_k
  \left(\ah_{ph,\kk}+\ahd_{ph,-\kk}\right)\left(\ah_{ph,-\kk}+\ahd_{ph,\kk}\right)\nonumber  \\ 
&\!+\!& \frac{i \hbar C_k \Omega_C}{\omega_e}
\left(\ah_{m,\kk}-\ahd_{m,-\kk}\right)
\left(\ah_{ph,-\kk}+\ahd_{ph,\kk} \right). \eqname{H_int}
\end{eqnarray}
All terms consist of quadratic forms in the creation and annihilation operators of the electromagnetic or the matter polarization fields. 
The first term $H_{ph,\kk}$ is the free e.m. field Hamiltonian. 
The second term $H_{at,\kk}$ describes the internal dynamics of the dressed atoms. 
The three lines of the third term $H_{int,\kk}$ respectively account for (i) the dipole coupling of the $g\rightarrow e$ transition to the e.m. field, (ii) the photon renormalization due to the squared vector potential term, and (iii) the coupling between the photon quantum field and the $m$ excitation as a result of the dressing field
\footnote{Though small, this term is crucial to preserve gauge invariance and avoid unphysical behaviors in the $k\rightarrow 0$ limit.}.

The coupling constant $C_k$ is equal to
\begin{equation}
C_k=\sqrt{\frac{2\pi \omega_e^2 n_{at}}{\hbar c k}}\,d_{eg},
\end{equation}
and $D_k=C_k^2/\omega_e$. As expected, these expressions involve the model parameters $q$ and $M$ only via their physical combination $d_{eg}$.

Introducing the vector ${\alphah}_\kk$ of bosonic operators: 
\begin{equation}
{\alphah}_\kk=(\ah_{ph,\kk},\ah_{e,\kk},\ah_{m,\kk},
\ahd_{ph,-\kk},\ahd_{e,-\kk},\ahd_{m,-\kk})^T,
\eqname{vec_op}
\end{equation}
and the Bogoliubov metric $\eta=\textrm{diag}[1,1,1,-1,-1,-1]$,
the Hamiltonian \eq{H_tot} can be recast in a simple matricial form:
\begin{equation}
H=\frac{\hbar}{2}\sum_{\kk} \alphah_{\kk}^{\dagger}\, \eta\,{\mc H}_\kk \alphah_{\kk}+E_0
\eqname{Ham}
\end{equation}
in terms of a $6\times6$ Hamiltonian matrix ${\mc H}_\kk$ of the form:
\begin{equation}
{\mc H}_\kk=
\left(
\begin{array}{cc}
{\mc K}_\kk & {\mc Q}_\kk \\
-{\mc Q}_\kk^\dagger & -{\mc K}_\kk^T
\end{array}
\right).
\eqname{H_matrix}
\end{equation}
where ${\mc K}_k$ and ${\mc Q}_k$ are $3\times3$ matrices. The constant $E_0$ fixes the energy zero: as it has no consequences in what follows, it will be neglected from now on.

The Hermitian matrix
\begin{equation}
{\mc K}_\kk=
\left(
\begin{array}{ccc}
c k+2D_k & iC_k & i\Omega_C C_k/\omega_e \\
-iC_k & \omega_e & \Omega_C \\
-i\Omega_C C_k/\omega_e & \Omega_C & \tilde{\omega}_m
\end{array}
\right)
\end{equation}
takes into account the free field, the internal atomic dynamics including the dressing beam, as well as the light-matter interaction terms at the level of the so-called Rotating Wave Approximation (RWA): whenever a radiative photon is absorbed (emitted), an atomic
excitation is created (destroyed) at its place~\cite{CCT4}.

The symmetric matrix
\begin{equation}
{\mc Q}_\kk=
\left(
\begin{array}{ccc}
2D_k & -iC_k & -i\Omega_C C_k/\omega_e \\
-iC_k & 0 & 0 \\
-i\Omega_C C_k/\omega_e & 0 & 0
\end{array}
\right)
\end{equation}
corresponds instead to those additional terms which describe anti-RWA, off-shell processes where a photon and an atomic excitation are simultaneously destroyed or created. 

The relative importance of the RWA $\mc{K}_\kk$ and the anti-RWA $\mc{Q}_\kk$ terms is quantified by the ratio $\bar{C}/\omega_e$ of the radiation-matter coupling strength $\bar{C}=C_{k=\omega_e/c}$ and the excitation frequency $\omega_e$.
For most atomic systems of actual experimental interest, this parameter is generally quite small: as a simplest example, consider the $D_2$ line of $^{87}$Rb atoms at $\lambda_e=2\pi\,c/\omega_e\simeq780\,\textrm{nm}$. 
The electric dipole moment of the transition is $d_{eg}\simeq 4.2\,e\,a_{Bohr}$~\cite{steck} and a typical value of lattice spacing is $a_L=300\,\textrm{nm}$.
For a unit filling factor $n=1$, the light matter coupling parameter is then $\bar{C}/\omega_e \simeq 1.7\cdot 10^{-4}$.
Although the condition $\bar{C}/\omega_e \ll 1$ rules out the possibility of observing the so-called ultra-strong coupling regime~\cite{ISB-QW-th,ISB-QW-th_PRL} in such dilute atomic systems, the anti-RWA terms in the Hamiltonian can still have interesting observable consequences as we shall see in what follows.

\section{Stationary state: ground state and polariton excitations}
\label{sec:stationary}

We begin the study of the optical properties of the system from the simplest case where the dressing parameters $\omega_C$ and $\Omega_C$ are kept fixed in time. For each value of them, the quadratic structure of the Hopfield Hamiltonian \eq{Ham} guarantees that this can be set into the canonical form:
\begin{equation}
H=\sum_{\kk,r} \hbar\,\omega_{r,\kk}
\,\phd_{r,\kk}\, \ph_{r,\kk}+E'_0
\eqname{H_quad}
\end{equation} 
by means of a Hopfield-Bogoliubov transformation~\cite{Hopfield}.
As in \eq{Ham}, the constant $E'_0$ is the zero-point energy and will be neglected from now on.
For each wavevector $\kk$, the frequencies $\omega_{r,\kk}$ of the elementary excitations are given by the eigenvalues corresponding to the positive-norm eigenvectors of the Hopfield-Bogoliubov matrix
\begin{equation}
{\mc M}_\kk=\left({\mc H}_\kk\right)^T.
\eqname{M}
\end{equation}
In the system under consideration here, the elementary excitations are the {\em lower} ($r=LP$), {\em middle} ($r=MP$, often also called {\em dark-state polariton}, e.g. in~\cite{fleischhauer}) and {\em upper} ($r=UP$) {\em polariton} modes. All these modes are linear superposition of light and matter excitations. The polaritonic annihilation operators $\ph_{r,k}$ can be written in terms of the eigenvectors $\vec{w}_{r,k}$ as:
\begin{equation}
\ph_{r,\kk}=
\sum_{\lambda=1}^{6}\,w_{r,\kk}^{\lambda}\,\alphah_\kk^{\lambda}.
%w_{q,\kk}\al{1}\,\ah_{ph,\kk}+
%w_{q,\kk}\al{2}\,\ah_{e,\kk}+
%w_{q,\kk}\al{3}\,\ah_{m,\kk}+
%w_{q,\kk}\al{4}\,\ahd_{ph,\kk}+ \\
%+w_{q,\kk}\al{5}\,\ahd_{e,\kk}+
%w_{q,\kk}\al{6}\,\ahd_{m,\kk},
\eqname{eig_vect}
\end{equation}
The index $\lambda$ runs over the six components of the eigenvector $\vec{w}_{r,k}$ of the Hopfield-Bogoliubov matrix \eq{M} and of the operator vector $\alphah_\kk$ defined in \eq{vec_op}.
An analogous expression holds for the creation operators $\phd_{r,\kk}$.
% The orthonormality condition of the eigenvectors in the Bogoliubov metric $\eta$
% \begin{equation}
% \vec{w}_{q,\kk}^\dagger\,\eta\,\vec{w}_{q',\kk}=\delta_{qq'}\,\varepsilon_q
% \eqname{normal}
% \end{equation}
% guarantees that the polaritonic operators $\ph_{q,\kk}$ satisfy the standard Bose commutation rules. The sign $\varepsilon_q$ of the eigenvector norm is $1$ for $q=1\ldots 3$ and $-1$ for $q=4\ldots 6$.

Grouping the $\ph_{r,\kk}$'s in the operator vector
\begin{equation}
\hat{\pi}_\kk=(
\ph_{LP,\kk},\ph_{MP,\kk},\ph_{UP,\kk},
\phd_{LP,-\kk},\phd_{MP,-\kk}\phd_{UP,-\kk})^T,
\end{equation}
the transformation \eq{eig_vect} to the polaritonic basis can be cast in the simple matricial
form $\hat{\pi}_\kk=\mathbf{W}_\kk\,\alphah_\kk$; for instance, the column vectors $\vec{\textrm{w}}_{\lambda}$ obtained as the transposed of the first three lines of $\mathbf{W}_\kk$ correspond to the $\vec{w}_{r,\kk}$ eigenvectors \eq{eig_vect}, $\vec{\textrm{w}}_{\{1,2,3\},\kk}=\vec{w}_{\{LP,MP,UP\},\kk}$.

The orthonormality condition of the eigenvectors in the Bogoliubov metric $\eta$ corresponds to the $\eta$-unitary condition
\begin{equation}
\mathbf{W}_\kk^{-1}=\eta\,\mathbf{W}_\kk^\dagger\,\eta,
\end{equation}
and guarantees that the operators $\ph_{q,\kk}$ satisfy the standard Bose commutation rules. 
In the $\hat{\pi}$ basis, the Hamiltonian matrix has the simple
diagonal form:
\begin{multline}
{\mc H}'_\kk=\mathbf{W}_\kk {\mc H}_\kk\,\mathbf{W}_\kk^{-1}=
\textrm{diag}[
\omega_{LP,k},\omega_{MP,k},\omega_{UP,k},\\
-\omega_{LP,k},-\omega_{MP,k},-\omega_{UP,k}].
\end{multline}

In view of the following developments, it is useful to give the explicit form of the $r=\{LP,MP,UP\}$ polariton operators in terms of the photonic ($ph$) and matter ($e,m$) excitation ones:
\begin{multline}
\ph_{r,\kk}=u^{ph}_{r,\kk}\,\ah_{ph,\kk}+u^{e}_{r,\kk}\,\ah_{e,\kk}+
u^{m}_{r,\kk}\,\ah_{m,\kk} \\
+v^{ph}_{r,\kk}\,\ahd_{ph,-\kk}+v^{e}_{r,\kk}\,\ahd_{e,-\kk}+
v^{m}_{r,\kk}\,\ahd_{m,-\kk}
\eqname{a->p}
\end{multline}
as well as the inverse transformation ($j=\{ph,e,m\}$):
\begin{multline}
\ah_{j,\kk}=u^{j*}_{LP,\kk}\,\ph_{LP,\kk}+u^{j*}_{MP,\kk}\,\ph_{MP,\kk}+
u^{j*}_{UP,\kk}\,\ph_{UP,\kk} \\
-v^{j}_{LP,\kk}\,\phd_{LP,-\kk}-v^{j}_{MP,\kk}\,\phd_{MP,-\kk}-
v^{j}_{UP,\kk}\,\phd_{UP,-\kk}.
\eqname{p->a}
\end{multline}
The $u$ and $v$ Hopfield coefficients characterize respectively the normal and anomalous weights of the different $ph,e,m$ components of the $LP,MP,UP$ polaritons.

\subsection{The polariton vacuum}
\label{sec:ground}

The vacuum state of the system corresponds to the ground state $|{\rm G}\rangle$  of the Hamiltonian \eq{H_quad}, and is defined by the vacuum condition
\begin{equation}
\ph_{r,\kk}\,|{\rm G}\rangle=0
\eqname{ground}
\end{equation}
for all polariton modes $r=\{LP,MP,UP\}$.

As both annihilation and creation operators are involved in the Bogoliubov transformation \eq{p->a}, the ground state $|{\rm G}\rangle$ corresponds in the original $\ah_{(ph,e,m)}$ basis to a squeezed vacuum state with a non-vanishing  expectation value of the photon and atomic excitation numbers:
\begin{multline}
N_{(ph,e,m)}^G=\langle G |  \ahd_{(ph,e,m),\kk} \ah_{(ph,e,m),\kk} | G \rangle = \\  =\sum_{r=\{LP,MP,UP\}}\,|v_{r,\kk}^{(ph,e,m)}|^2.
\eqname{N_v}
\end{multline}
As the atomic systems under consideration here are far from the ultra-strong coupling regime~\cite{ISB-QW-th,ISB-QW-th_PRL,ISB-QW-exp}, an accurate estimation of $N_{(ph,e,m)}^G$ can be obtained by means of perturbation theory in the light-matter coupling strength $\bar{C}/\omega_e \ll 1$. The dressing amplitude $\Omega_C$ is assumed to be at most of the order of $\bar{C}$.

The zeroth order approximation $\vec{w}^{0}_{q,\kk}$ of the eigenvector can be obtained by diagonalizing the block diagonal zeroth order Hopfield-Bogoliubov matrix
\begin{equation}
{\mc M}^0_\kk=
\left(
\begin{array}{cc}
{\mc K}^T_\kk & 0 \\
0 & -{\mc K}_\kk
\end{array}
\right).
\eqname{M_matrix_0}
\end{equation}
This provides the zeroth order eigenvalues $\omega^0_{\lambda,\kk}$ and eigenvectors $\vec{{\textrm w}}^{0}_{\lambda,\kk}$.
These latter have the form
\begin{eqnarray}
\vec{{\textrm w}}^{0}_{\lambda,\kk}&\!\!\!=\!\!\!&(u^{ph,0}_{\lambda,\kk},u^{e,0}_{\lambda,\kk},u^{m,0}_{\lambda,\kk},0,0,0)^T\hspace{0.4cm} (\lambda=1,2,3), \\
\vec{{\textrm w}}^0_{\lambda,\kk}&\!\!\!=\!\!\!&(0,0,0,u^{ph,0*}_{\lambda-3,\kk},u^{e,0*}_{\lambda-3,\kk},u^{m,0*}_{\lambda-3,\kk})^T\hspace{0.cm} (\lambda=4,5,6). \nonumber \\
&&
\end{eqnarray}
Up to this level of approximation the virtual occupation \eq{N_v} is then rigorously vanishing.

When evaluating the first order correction, attention has to be paid to the non-positive nature of the $\eta$-metric:
\begin{equation}
\vec{{\textrm w}}^{1}_{\lambda,\kk}=\sum_{\stackrel{\lambda'=\{1\dots6\}}{\lambda'\neq \lambda}}\varepsilon_{\lambda'}\,\frac{\vec{{\textrm w}}^{0\,\dagger}_{\lambda',\kk}\,\eta\;\delta{\mc M}_k\,\vec{{\textrm w}}^{0}_{\lambda,\kk}}{\omega^0_{\lambda,\kk}-\omega^0_{\lambda',\kk}}\,\vec{{\textrm w}}^0_{\lambda',\kk},
\eqname{w1}
\end{equation}
where the sign $\varepsilon_\lambda$ is $+1$ for $\lambda=1,2,3$ and $-1$ for $\lambda=4,5,6$.
The perturbation matrix $\delta{\mc M}_k={\mc M}_k-{\mc M}_k^0$ is of order $\bar{C}/\omega_e$ and has non-zero entries only in the off-diagonal $3\times 3$ blocks.
Keeping in \eq{w1} only the lowest order terms in $\bar{C}/\omega_e$, one gets to the first order corrections:
\begin{equation}
v^{j,1}_{r,\kk}=\sum_{r'} \frac{-i\,C_k}{\omega^0_{r,\kk}+\omega^0_{r',\kk}}\,
(u_{r',\kk}^{ph,0}\,u_{r,\kk}^{e,0}+u_{r',\kk}^{e,0}\,u_{r,\kk}^{ph,0}   )\,u^{j,0*}_{r',\kk},
\eqname{v1}
\end{equation}
where the sum runs over the three polariton branches $r'=\{LP,MP,UP\}$.
Note that the off-diagonal terms due to the squared vector potential give a contribution to \eq{v1} whose amplitude is of the order of $D_k/\omega_e = (C_k/\omega_e)^2$ and have been therefore neglected. The same for the off-diagonal terms due to the dressing field, whose contribution is of the order of $C_k \Omega_C/\omega_e^2$.

The virtual population in the ground state is then of the order of $(C_k/\omega_e)^2$: exception made for a small region around $\kk=0$ where $C_k$ diverges, this population is therefore very small for all polariton branches. This fact provides an {\em a posteriori} justification of the use of perturbation theory.

\begin{figure}
\begin{center}
\includegraphics[width=\columnwidth,clip]{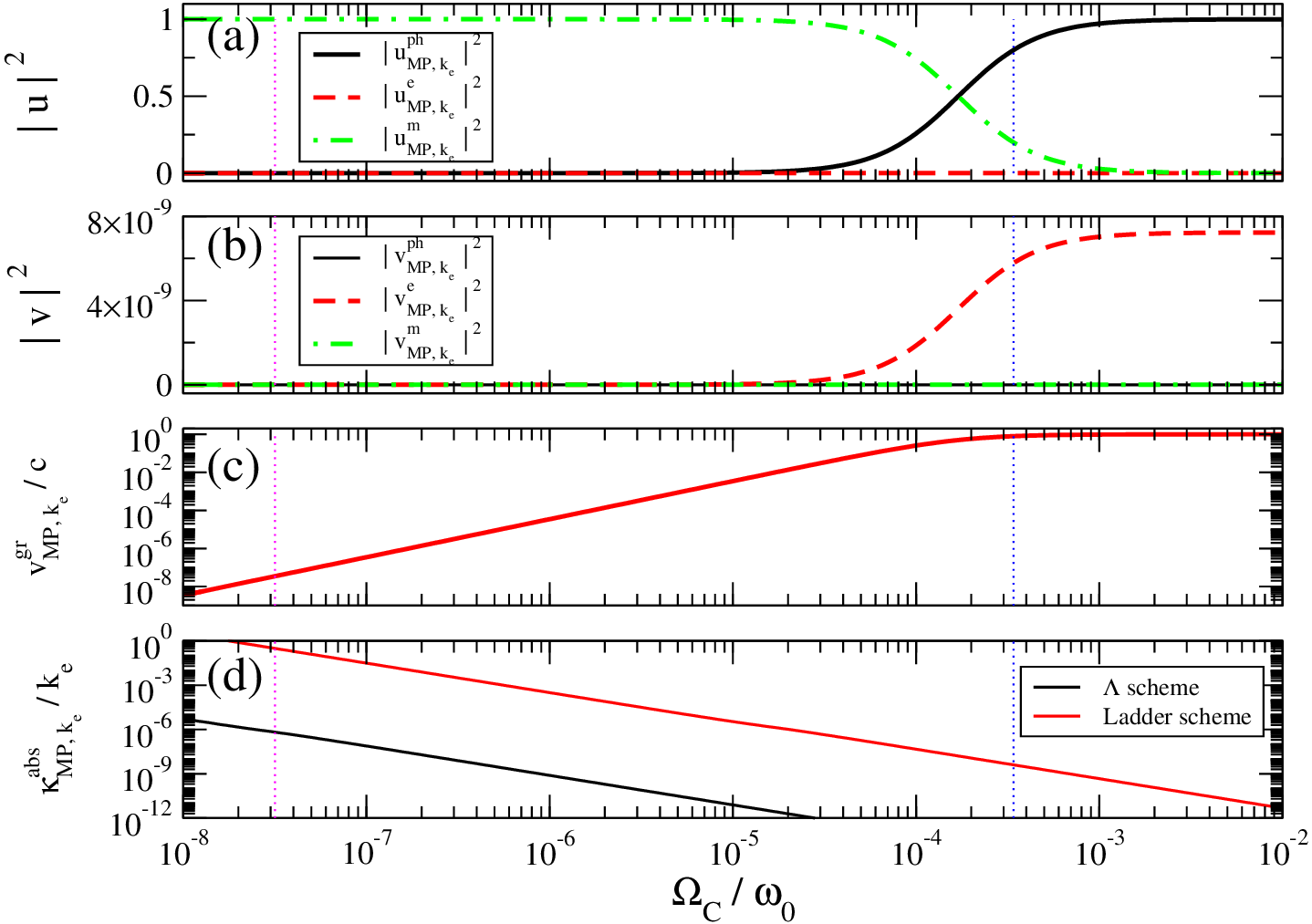}
\end{center}
\caption{Hopfield $u$ and $v$ coefficients (a,b), the group velocity (c) and the absorption coefficient (d) of the middle polariton (MP) as a function of dressing amplitude $\Omega_C$ in the resonant ($\omega_e=\tilde{\omega}_m$) case.
On the scale of the figure, the analytical approximations \eq{vgr}, (\ref{eq:u0_ph}-\ref{eq:u0_m}) and \eq{v_e_fin} are undistinguishable from the exact calculations. Vertical lines indicate the dressing amplitude values used in the next figures.
The system parameters correspond to the $D_2$ line of a Mott insulator system of $^{87}$Rb atoms with filling factor $n=1$ in a $a_L=300\,$nm lattice: $\bar{C}/\omega_e=1.7\cdot 10^{-4}$, $\lambda_e=2\pi/k_e=2\pi\,c/\omega_e=780\,\textrm{nm}$.
}
\label{fig:MP_summ}
\end{figure}
In the most significant resonance region $k\simeq k_e=\omega_e/c$, the frequency denominator of \eq{v1} can be approximated by $2\omega_e$ in a sort of degenerate polariton approximation:
\begin{eqnarray}
v^{ph,1}_{r,k}&\simeq& -\frac{i\,\bar{C}}{2\omega_e}\,u^{e,0}_{r,k}, \eqname{v_p} \\
v^{e,1}_{r,k}&\simeq& -\frac{i\,\bar{C}}{2\omega_e}\,u^{ph,0}_{r,k}, \eqname{v_e} \\
v^{m,1}_{r,k}&\simeq& 0 \eqname{v_m}.
\end{eqnarray}
Compact formulas for the fully resonant point $\omega_e=\tilde{\omega}_m$, $k=k_e$ of the MP are immediately obtained by inserting in (\ref{eq:v_p}-\ref{eq:v_m}) the explicit form of the RWA eigenvectors of the zeroth order Hopfield-Bogoliubov matrix \eq{M_matrix_0}:
\begin{eqnarray}
 u^{ph,0}_{MP,k_e}&=& \frac{\Omega_C}{(\bar{C}^2+\Omega_C^2)^{1/2}}, \eqname{u0_ph} \\
 u^{e,0}_{MP,k_e}&=& 0 \eqname{u0_e}, \\
 u^{m,0}_{MP,k_e}&=& -\frac{i\,\bar{C}}{(\bar{C}^2+\Omega_C^2)^{1/2}}, \eqname{u0_m}
\end{eqnarray}
which leads to anomalous amplitudes
\begin{eqnarray}
v^{ph,1}_{MP,k_e}&\simeq&v^{m,1}_{MP,k_e}\simeq0, \\
v^{e,1}_{MP,k_e}&\simeq& -\frac{i\,\bar{C}\,\Omega_C}{2\omega_e(\bar{C}^2+\Omega_C^2)^{1/2}}. \eqname{v_e_fin}
\end{eqnarray}
The accurateness of these analytical approximations is visible in Fig.\ref{fig:MP_summ}(a,b) where the Hopfield $u$ and $v$ coefficients are plotted for the fully resonant $k=k_e$ point of the MP as a function of the dressing amplitude: on the scale of the figure the analytical approximations are undistinguishable from the exact calculations.

As long as the system parameters are kept constant in time, the virtual populations \eq{N_v} are intrinsecally bound to the system ground state and can not be revealed by a standard photodetector based on absorption processes~\cite{ISB-QW-th,ISB-QW-th_PRL}.
On the other hand, the dependence of the anomalous amplitude \eq{v_e_fin} on $\Omega_C$ [see Fig.\ref{fig:MP_summ}(b)] suggests that the zero-point fluctuations in the ground state can be externally controlled by varying $\Omega_C$ in time.
In particular, if the time-modulation of $\Omega_C$ is sufficiently fast, the system is not able to adiabatically follow the time-dependent vacuum state. 
As a consequence, real polaritons are created in the system and then emitted as radiative photons into the surrounding free space where they can be detected by a photodetector. This will be the subject of the next Secs.\ref{sec:time-dep}-\ref{sec:quant}.

\subsection{Polariton dispersion}

\begin{figure*}
\begin{center}
\includegraphics[width=0.9\textwidth,clip]{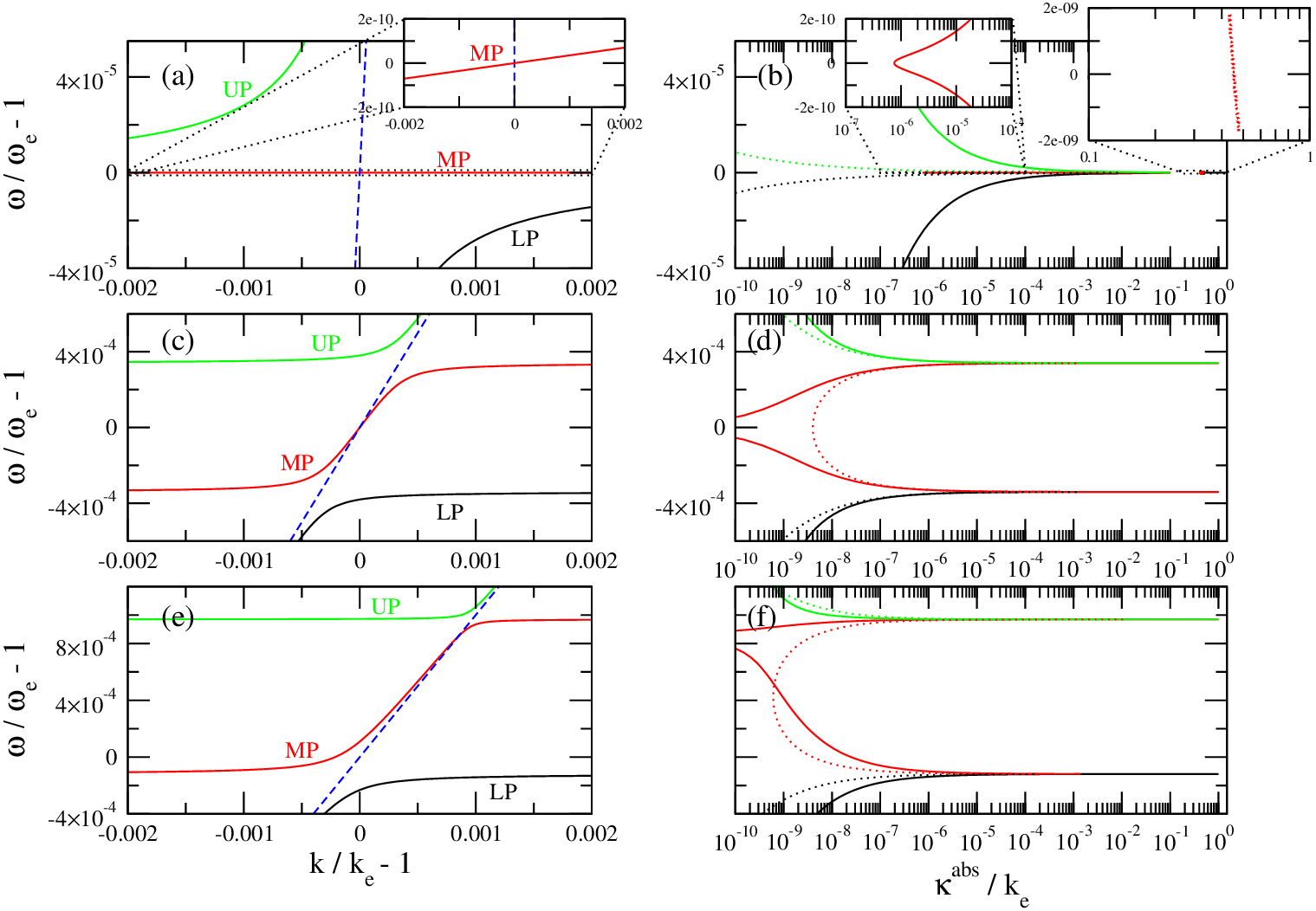}
\end{center}
\caption{Dispersion relation (a,c,e) and absorption coefficient (b,d,f) of the three polariton modes. Same system parameters as in Fig.\ref{fig:MP_summ}.
Black, red and green lines refer to the lower polariton (LP), the middle polariton (MP), the upper polariton (UP), respectively.
The blue dashed line in (a,c,e) is the free photon dispersion $\omega=c k$.
Panels (a,b): resonant $\omega_e=\tilde{\omega}_m$, weak dressing $\Omega_C/{\bar C}= 1.86\cdot 10^{-4}\ll 1$ case. Insets in (a,b): magnified views of the MP.
Panels (c,d): resonant $\omega_e=\tilde{\omega}_m$, strong dressing $\Omega_C/{\bar C}=2$ case.
Panels (e,f): non-resonant $\tilde{\omega}_m-\omega_e=5\,\bar{C}$, strong dressing $\Omega_C/{\bar C}=2$ case.
In the absorption (b,d,f) panels: $\Lambda$ level scheme with $\gamma_e/2\pi=6\,\textrm{MHz}$ and $\gamma_m/2\pi=10\,\textrm{Hz}$ (solid lines); ladder level scheme with exchanged $\gamma_e/2\pi=10\,\textrm{Hz}$ and $\gamma_m/2\pi=6\,\textrm{MHz}$ values (dotted lines).
}
\label{fig:polar_disp}
\end{figure*}

Examples of polariton dispersion are shown in Fig.\ref{fig:polar_disp} for the most significant cases. The shape of the three polariton branches changes in a substantial way depending on the dressing parameters $\omega_C$ and $\Omega_C$, which provides a simple way to externally vary the optical properties of the system in real time. 
An application of atomic Mott insulators as dynamic photonic structures was indeed proposed in~\cite{bariani}.

As long as linear optical properties are considered, it is important to note that the predictions of the quantum model are indistinguishable from the solution of the Maxwell equations including the semiclassical expression for the local dielectric polarizability of three-level atoms~\cite{Arimondo}.
In particular, this is the case of the band dispersions shown in Fig.\ref{fig:polar_disp}.

As $\omega_e$ and $\tilde{\omega}_m$ have no spatial dispersion, the three bands have no spectral overlap and are separated by energy gaps in which the radiation can not propagate. 
For the systems under consideration here, the gaps between the bands are however very narrow, on the order of $\bar{C}^2/\omega_e$, and therefore almost invisible on the scale of the figure.

When the dressing is on resonance with the $m\rightarrow e$ transition ($\omega_e=\tilde{\omega}_m$), the atomic resonance is split into a symmetric Autler-Townes doublet at $\omega_e\pm \Omega_C$ and the photonic mode anticrosses each of the two components. 
Two sub-cases are to be distinguished: for a strong dressing $\Omega_C\gtrsim{\bar C}$ [panel (c)], the two anticrossings are almost completely separated and have a half-width equal to ${\bar C}/\sqrt{2}$; in between the two anticrossings (i.e. in the neighborhood of $\omega_e$), the middle polariton (MP) almost coincides with the photon branch and has a group velocity ${\textrm v}^{gr}_{MP,k}=d\omega_{MP,k}/dk$
close to $c$.
On the other hand, for a weak dressing $\Omega_C\ll{\bar C}$, the two anticrossings overlap, which results in a strong mutual distortion [panel (a)]: the MP dispersion is strongly flattened, and its group velocity becomes orders of magnitude slower than $c$ [inset of panel (a)].

An approximate, yet quantitatively very accurate expression for the group velocity ${\rm v}^{gr}_{MP,k_e}$ of the MP around resonance $k=k_e=\omega_e/c$ is easily obtained from the MP eigenvector (\ref{eq:u0_ph}-\ref{eq:u0_m}) of the RWA Hopfield-Bogoliubov matrix \eq{M_matrix_0}:
\begin{equation}
{\rm v}^{gr}_{MP,k_e}=c\,|u_{MP,k_e}^{ph,0}|^2= c\,\frac{\Omega_C^2}{\Omega_C^2+{\bar C}^2}.
\eqname{vgr}
\end{equation}
As one can see in Fig.\ref{fig:MP_summ}(a), arbitrarily slow values of ${\rm v}^{gr}_{MP,k_e}$ can be obtained by simply reducing the amplitude $\Omega_C$ of the dressing field: thanks to the trapping of atoms at lattice sites, no lower bound to the group velocity appears as a consequence of the atomic recoil in absorbing and emitting photons~\cite{JETP}.
For the specific case of Rb atoms considered in the figures, a quite conservative value $\Omega_C/2\pi=12\,\textrm{MHz}$ (i.e. twice the radiative lifetime of the $e$ state) already leads to ${\rm v}^{gr}_{M,k_e}=11\,\textrm{m/s}$.

In the case of an non resonant dressing [panel (e)], the oscillator strength of the optical transition is shared in an asymmetric way by the two components of the Autler-Townes doublet. This fact is responsible for the anticrossing to be wider for the component having a larger $e$ state weight (the lower one in the figure). Because of the optical Stark effect, the position of this main anticrossing is slightly shifted by $\Delta\omega_{OSE}=\Omega_C^2/(\omega_e-\tilde{\omega}_m)$ from its bare position at $\omega_e$.

\subsection{Effect of losses}
\label{sec:losses}

Losses from both the $e$ and the $m$ states are responsible for the finite lifetime of polaritonic excitations. For typical systems, both decay rates $\gamma_{(e,m)}$ are much smaller than the radiation-matter coupling $\bar{C}$ and the energy splitting $\omega_{q,k}-\omega_{q',k}$ between different polariton bands $q\neq q'$ at a given wave vector $\kk$.
In optics, this regime goes often under the name of {\em strong coupling} regime~\cite{CQED,Microcav} and is characterized by losses not being able to effectively mix the different polariton branches, which retain their individuality.

The Fermi golden rule then provides an accurate prediction for the decay rate $\gamma_{r,\kk}$ of the plane wave polariton state of wave vector $\kk$ on the $r$ branch: 
\begin{equation}
\gamma_{r,\kk}=\gamma_e\,|u^e_{r,\kk}|^2+\gamma_m\,|u^m_{r,\kk}|^2;
\eqname{gamma_q}
\end{equation}
physically, this decay rate is the relevant one in a light stopping experiment where light is stored in the system during a macroscopically long time~\cite{stop,fleischhauer}.
In a propagation configuration (i.e. at given frequency), the relevant quantity is instead the absorption length~\cite{tait}:
\begin{equation}
\ell_{r,\kk}=\left(\kappa^{abs}_{r,\kk}\right)^{-1}={\rm v}^{gr}_{r,\kk}/\gamma_{r,\kk}.
\eqname{ell_q}
\end{equation}
The relative value of the $\gamma_{(e,m)}$ loss rates depends on the specific level scheme under consideration. 

In a $\Lambda$ configuration, $\gamma_m$ is mostly given by non-radiative effects, and generally has a very small value. As no intrinsic effect is expected to significantly contribute to $\gamma_m$ in Mott insulator states, it can in principle be suppressed to arbitrarily small values by means of a careful experimental setup.
On the other hand, radiative decay from the $e$ state into the $m$ state can take place by spontaneous emission of a photon into a mode different from the coherent dressing one. The rate $\gamma_e$ of this process is equal to the one of an isolated atom in free space
\footnote{In our formalism, such processes correspond to the conversion of a $e$ excitation into a $m$ excitation plus a polariton at a frequency $\omega_e-\omega_m$ far from the resonance region $\omega_e-\omega_g$.
Using Fermi golden rule, it is immediate to see that the corresponding decay rate is indeed equal to the free space spontaneous radiative decay rate for the $e\rightarrow m$ transition.}, 
and can be reduced only by choosing an optically weak $m\rightarrow e$ transition to dress the system.

In a ladder configuration, the roles of $\gamma_{e}$ and $\gamma_m$ are exchanged. Spontaneous emission processes can now occur on the $m\rightarrow e$ transition giving rise to a significant $\gamma_m$ value.
On the other hand, provided the $g\rightarrow e$ transition is closed and no other decay channel is available for the $e$ state, no spontaneous process can contribute to $\gamma_e$, which can then be made arbitrarily small by means of a careful design of the experimental set-up.

This remarkable fact was discussed at length in the pioneering paper~\cite{Hopfield}, where it was pointed out that polaritons in a lattice of two-level atoms are not subject to dissipation and can propagate with a slow group velocity along macroscopic distances.
This fact originates from the rigidity of the lattice that freezes the atomic motion in the ground state of each site and prevent atoms from incoherently scattering the light.
As recently suggested in~\cite{MI_optics,Zoubi}, atomic Mott insulators then constitue an ideal realization of this two-level Hopfield model.

Plots of the prediction \eq{ell_q} for the absorption coefficient for the three polariton branches are shown in Figs.\ref{fig:polar_disp}(b,d,f) for respectively the $\Lambda$ (solid lines) and the ladder (dashed lines) three-level schemes~\footnote{Note that this plot does not include the extinction effects that occur for frequencies within the (very small) gaps between the bands, and which are e.g. responsible for the operation of distributed Bragg reflectors~\cite{Burstein,macleod}.}.
Again, these curves are in perfect agreement with the solution of the Maxwell equations using the semiclassical expression for the dielectric polarizability of three-level atoms~\cite{Arimondo}.
As expected, absorption is strongly peaked in the anticrossing regions where polaritonic bands have the largest weight of matter excitations and the slowest group velocity. This, in both cases of $\Lambda$ and ladder configuration.

The only exception is the dip that is visible exactly on resonance with the two-photon Raman transition $\omega=\tilde{\omega}_m$ in the case of a $\Lambda$ configuration.
In optics, this effect goes under the name of Electromagnetic Induced Transparency (EIT) effect~\cite{Arimondo,EIT_fleisch}: in the vicinity of the resonance, quantum interference suppresses the weight of $e$ excitation in the $MP$ mode. Consequently, the absorption rate is quenched to the non-radiative one $\gamma_{MP,k_e}\simeq \gamma_m\ll\gamma_e$. 
This effect is most dramatic in the case of a weak dressing $\Omega_C\ll\bar{C}$ [see in particular the left inset of Fig.\ref{fig:polar_disp}(b)], where the minimum of the spatial absorption coefficient $\kappa^{abs}_{MP,k_e}$ around resonance remains remarkably deep in spite of the very slow group velocity ${\rm v}^{gr}_{MP,k_e}\ll c$.

This allows for MP polaritons to propagate for macroscopic distances without being appreciably absorbed~\footnote{For the Rb atoms under consideration here, an absorption length of $1\,\textrm{m}$ corresponds to $\kappa^{abs}/k_e=1.24\cdot10^{-7}$.}: 
as one can see in Fig.\ref{fig:MP_summ}(b), a group velocity as slow as ${\rm v}^{gr}_{MP,k_e}=15\,\textrm{m/s}$ still corresponds to an absorption length as large as $\ell_{MP,k_e}\simeq 16\,\textrm{cm}$.
\\

To conclude the section, it is useful to summarize the main advantages of using a Mott insulator for slow-light experiments:

\begin{enumerate}
\item As shown in the pioneering paper~\cite{Hopfield}, slow light propagation can be observed in a lattice of two-level atoms almost without absorption.

\item Systems of three-level atoms allow for a real-time external control of the optical properties (polariton dispersion, group velocity, zero-point fluctuations) by simply varying the frequency $\omega_C$ and the amplitude $\Omega_C$ of the dressing beam.

\item As the atoms interact with each other only via the electromagnetic field, no intrinsic decoherence effect can contribute to the $\gamma_m$ rate in a $\Lambda$ configuration nor to the $\gamma_e$ one in a ladder configuration.

\item The presence of a single atom at each lattice site eliminates the inhomogeneous broadening of the transitions that originates e.g. from the spatially varying density profile of a Bose-Einstein condensate.

\item The trapping of atoms at the lattice sites eliminates the lower bound to the group velocity that atomic recoil would impose in the case of a homogeneous, untrapped gas~\cite{JETP}.

\end{enumerate}

\section{Dynamical Casimir emission in the presence of a time modulation}
\label{sec:time-dep}

When the dressing parameters are modulated in time, the quadratic form of the Hamiltonian is preserved, yet with a time-depedent Hamiltonian matrix $\mc{H}_\kk(t)=\bar{\mc H}_\kk+\delta\mc{H}_\kk(t)$.
At each time $t$, a Bogoliubov transformation diagonalizing the istantaneous Hamiltonian can be still found, but the transformation matrix $\mathbf{W}_\kk(t)$, as well as the polariton bands $\omega_{r,k}$ and the expression of the polariton operators $\ph_{r,\kk}$ in terms of the original $\ah_{j,\kk}$ ones are now varying in time, as well as the vacuum state $|G(t)\rangle$ of the system.
While for slow modulations the system is able to adiabatically follow the istantaneous ground state $|G(t)\rangle$, excitations are created in the case of a faster modulation. The study of the properties and the intensity of this dynamical Casimir emission is the subject of the present and the next sections. 

Our strategy is to reduce the problem to a simple and tractable parametric Hamiltonian to which the standard tools of quantum optics can be applied.
In particular, we shall concentrate our attention on the simplest case of a weak time-modulation $\|\delta {\mc H}_\kk \| \ll \|\bar{\mc{H}}_\kk\|$ for which perturbation theory can be used to obtain analytical predictions: most among the perturbations that one can envisage to apply to the atomic system largely fulfill in fact this condition. A discussion of the physics beyond perturbation theory can be found in~\cite{lambrecht,dezael,ISB-QW-th,ISB-QW-th_PRL}

Let $\bar{\mathbf W}_\kk$ be the Bogoliubov transformation diagonalizing the unperturbed Hamiltonian matrix $\bar{\mc H}_\kk$. In general, the perturbation Hamiltonian 
\begin{equation}
\delta {\mc H}'_\kk(t)=\bar{\mathbf W}_\kk\,\delta{\mc H}_\kk(t)\,\bar{\mathbf W}_\kk^{-1}
\end{equation} 
is not diagonal in this basis.
In terms of polariton creation and annihilation operators, $\delta {\mc H}'_\kk$ introduces terms of two kinds~\cite{Law}. 
The diagonal $3\times3$ blocks corresponds to terms of the form $\phd_{r,\kk}\,\ph_{s,\kk}$, which are responsible for a renormalization of the polariton energies (for $r=s$) and for the occurrence of interbranch transitions (for $r\neq s$) which transfer already existing polaritons from one branch to another. At the lowest order, these terms have no effect on the polaritonic vacuum state and will therefore not be considered in what follows.

The off-diagonal $3\times3$ blocks of ${\mc H}'_\kk$ are more interesting in the present context, as they correspond to terms of the forms $\phd_{r,\kk}\,\phd_{s,-\kk}$ and $\ph_{r,\kk}\,\ph_{s,-\kk}$, which respectively create or destroy pairs of polaritons in the opposite $\pm\kk$ momentum states of the $r$ and $s$ branches ($r,s=\{LP,MP,UP\}$). In particular, they account for the creation of correlated pairs of polaritons out of the vacuum state of the system via parametric amplification of the zero-point quantum fluctuations. 

This emission of radiation is an example of the still unobserved dynamical Casimir effect (DCE)~\cite{DCE}.
The most celebrated example of DCE is predicted for a metallic cavity whose length is varied in time by means of a mechanical motion of its mirrors~\cite{lambrecht}. Another possibility consists of varying the effective length of the cavity by modulating the mirror conductivity~\cite{DCE_cond,PD}, by mimicking moving mirrors via a suitably chosen $\chi^{(2)}$ non-linear optical element~\cite{dezael}, or by varing the bulk refractive index of the cavity material~\cite{DCE_n,Law}.

A generalization of this latter scheme is the subject of the present paper: the modulation of the dielectric properties of the atomic medium is created by varying in time the dressing field amplitude. 
Using the formalism developed in the previous sections, we will be able to go beyond the non-dispersive dielectric approximation made by most of the existing works, and fully include the resonant dynamics of matter excitations. Using this microscopic model, accurate predictions will be obtained also for the frequency window in the neighborhood of the atomic resonance where light-matter coupling is the strongest as well as the expected intensity of the dynamical Casimir emission.

\subsection{Polariton emission rate in a bulk geometry}

Depending on the frequency spectrum of the time-modulation $\delta{\mc H}_\kk(t)$, 
polaritons can be emitted in any momentum state: a monochromatic oscillation at a frequency $\omega$ of the form 
$\delta{\mc H}_\kk(t)=\delta{\mc H}_\kk\,(e^{i\omega t}+e^{-i\omega t})$, is in fact able to resonantly create pairs of polaritons in the $r,s$ branches at wave vectors $\pm \kk$ fulfilling the parametric resonance condition
\begin{equation}
\omega_{r,\kk}+\omega_{s,-\kk}=\omega.
\eqname{resonant}
\end{equation}
Note that an analogous condition was recently obtained for the parametric emission of phonons in trapped atomic Bose-Einstein condensate in an optical lattice when the lattice potential is modulated in time~\cite{ParamPhonon}.

Starting from the polaritonic vacuum as initial state, and limiting ourselves to lowest-order effects in the modulation amplitude, we can safely neglect the $\phd_{r,\kk}\,\ph_{s,\kk}$ terms that arise from the modulation and rewrite the time-dependent system Hamiltonian \eq{Ham} in the standard parametric form~\cite{QO}: 
\begin{multline}
H={\bar H}+\delta H=\\
{\bar H}+\frac{\hbar}{2}\sum_{rs,\kk}
\left[V_{rs,\kk}\,e^{-i\omega t}\,\phd_{r,\kk}\,\phd_{s,-\kk}+\textrm{h.c.} \right].
\eqname{H_param}
\end{multline}
From this Hamiltonian the polariton creation rate is determined by means of the Fermi golden rule.
As final states, one has to consider pairs of polaritons created in the $r,s=\{LP,MP,UP\}$ branches at wavevectors $\pm\kk$:
\begin{equation}
 \frac{dN_{rs}}{dt}=\frac{2\pi}{\hbar^2}\,\sum_\kk 
\left|\langle r,\kk;s,-\kk|\delta H|G \rangle \right|^2\,\delta(\omega_{r,\kk}+\omega_{s,-\kk}-\omega).
\end{equation}
In the weak modulation regime (i.e. far below any parametric oscillation threshold), this approach is equivalent to other ones based e.g. on the input-output formalism; including all neglected terms is instead crucial if one is interested in the peculiar pulse shaping and frequency up-conversion effects discussed at length in~\cite{lambrecht,ISB-QW-th_PRL}.

Replacing as usual the the sum over $\kk$ vectors with an integral, the total creation rate per unit volume into the $r,s$ branches reads:
\begin{equation}
\frac{dN_{rs}}{dt\,d{\mc V}}=\frac{\bar{k}^2}{2\pi\,({\rm v}^{gr}_{r,\bar{k}}+{\rm v}^{gr}_{s,\bar{k}})}
\,|V_{rs,\bar{k}}|^2,
\eqname{rate}
\end{equation}
where the $r$ ($s$) polariton is assumed to be emitted in the $k_z>0$ ($k_z<0$) half-space.
$\bar{k}$ is the wavevector value at which the resonant condition \eq{resonant} is satisfied for the $r,s$ branches under examination. 
${\rm v}^{gr}_{(r,s),\bar{k}}$ are the group velocities, and the matrix element $V_{rs,k}$ of the process is given by: 
\begin{equation}
V_{rs,k}=\left.\bar{\mathbf{W}}_\kk\,\delta{\mc
  H}_\kk\,\bar{\mathbf W}_\kk^{-1}\right|_{r,s+3};
  \eqname{V_qrk}
\end{equation}
here, the $q,s=\{LP,MP,UP\}$ branches correspond to respectively $\{1,2,3\}$.
Perturbative expressions for the matrix element \eq{V_qrk} to the leading order in $\bar{C}/\omega_e$ will be given in the next subsection Sec.\ref{sec:matrix_element}.

It is worth noticing that a finite emission intensity $\langle \phd_{r,\kk} \ph_{r,\kk}\rangle \neq 0$ is obtained while the classical amplitudes remain strictly zero $p_{r,\kk}=\langle \ph_{r,\kk}\rangle=0$ during the time-modulation.
The dynamical Casimir emission is in fact a purely quantum effect due to the parametric amplification of the zero-point fluctuations of the polariton field.

Eqs.\eq{rate} and \eq{V_qrk} are the central result of the present section: they quantify the polariton emission in an idealized, spatially infinite system and will represent the central building block in the study of experimentally relevant finite-size geometries that we shall perform in Sec.\ref{sec:realistic}. A quantitative discussion of the actual value of the emission rate for realistic values of the system parameters will be given in Sec.\ref{sec:quant}.

\subsection{Approximate analytical expression of the matrix element}
\label{sec:matrix_element}

A simple, yet accurate estimation of the matrix element \eq{V_qrk} can be obtained by means of the perturbative approximation of the eigenvectors $\vec{w}_{r,k}$ discussed in sec.\ref{sec:ground}.
As a most significant example, we consider a periodic modulation of the coupling amplitude $\Omega_C$ of the form:
\begin{equation}
 \Omega_C(t)=\Omega_C+\delta \Omega_C\,\left(e^{-i\omega t}+e^{i\omega t}\right),
\eqname{Omega_C}
\end{equation}
while its frequency $\omega_C$ is kept constant
\footnote{A similar perturbative calculation for a periodic modulation of $\tilde{\omega}_m$ would lead to a vanishing result at the lowest order in perturbation theory. This is a direct consequence of the vanishing value of $v_{r,k}^{m,1}$ obtained in \eq{v_m}.
A complete calculation including next order terms results in a dynamical Casimir emission intensity orders of magnitude weaker than for a modulation of  $\Omega_C$.}.
Modulations of this kind were at the heart of recent light-stopping experiments~\cite{stop}, albeit on a much slower time-scale: in that case, the modulation had in fact to be adiabatic enough not to induce inter-branch transitions.

Inserting the perturbative result (\ref{eq:v_p}-\ref{eq:v_m}) into the $\bar{\mathbf{W}}_\kk$ transformation matrix, the following expression for the matrix element is found, which is valid throughout the whole resonance region:
%\begin{multline}
\begin{equation}
V_{rs,k}
%= \left[-u_{q,k}^e v_{r,k}^m - v_{q,k}^m u_{r,k}^e  - v_{q,k}^e u_{r,k}^m  - u_{q,k}^m %v_{r,k}^e \right..
%\\ \left. -\frac{i\bar{C}}{\omega_e} \left( u_{q,k}^{ph} u_{r,k}^m + u_{r,k}^{ph} %u_{q,k}^m   \right)\right]\,\delta \Omega_C \\ 
\simeq
- \frac{i\, \bar{C}}{2\omega_e}\,\left(u_{r,k}^{ph} u_{s,k}^m+u_{s,k}^{ph} u_{r,k}^m\right)\,\delta \Omega_C.
\eqname{analytic1}
%\end{multline}
\end{equation}
As expected, $V_{rs,k}$ is proportional to the amplitude of the dressing modulation and to $\bar{C}/2\omega_e$, i.e. the ratio between the amplitude of the antiresonant light-matter coupling and the energy associated to the creation of a pair of excitations. The presence in \eq{analytic1} of both photonic and $m$ excitation Hopfield coefficients means that a mixing between light and matter modes is necessary to obtain a sizeable emission. A most favourable region is therefore the fully resonant point where $\omega_e=\tilde{\omega}_m$ and the modulation is driven at a frequency $\omega=2\omega_{e}$.

An analytical estimation of $V_{rs,k_e}$ with $r=s=MP$ is readily obtained for this case by inserting in \eq{analytic1} the analytical eigenvector (\ref{eq:u0_ph}-\ref{eq:u0_m}) of the zeroth order Hopfield-Bogoliubov matrix \eq{M_matrix_0}:
\begin{equation}
V_{MP,k_e}=
\frac{\Omega_C}{\omega_e}\,\frac{\bar{C}^2}{\bar{C}^2+\Omega_C^2}\,
\delta\Omega_C.
\eqname{analytic}
\end{equation}
In the slow light regime $\Omega_C\ll \bar{C}$, the matrix element $V_{MP,k_e}$ grows as $\Omega_C\,\delta\Omega_C/\omega_e$, while it goes as $\bar{C}^2\,\delta\Omega_C/(\Omega_C\omega_e)$ for $\Omega_C\gg\bar{C}$.

Using the expression \eq{vgr} for the group velocity, it is immediate to see that for a given value of the relative modulation amplitude $\delta \Omega_C/\Omega_C$, the emission intensity \eq{rate} starts proportionally to $\Omega_C^2$ for small values $\Omega_C\ll \bar{C}$ and then saturates to a finite value for $\Omega_C\gg \bar{C}$. We will come back on these issues in the quantitative discussion of Sec.\ref{sec:quant}.\\

The main result \eq{analytic1} of the present section fully includes the dispersion of polaritons, which is crucial to correctly describe the region around resonance. This, in contrast with previous works~\cite{Law} where the Casimir emission was studied for a dispersionless dielectric medium with a periodically oscillating dielectric constant 
\begin{equation}
\epsilon(t)=\bar{\epsilon}+\delta\epsilon\,(e^{i\omega t}+e^{-i\omega t}).
\end{equation}
Also in this case, a Hamiltonian of the parametric form \eq{H_param} can however be written, with the matrix element
\begin{equation}
 V_{k}=\frac{\omega}{4\,\bar{\epsilon}}\,\delta\epsilon.
\eqname{V_k_diel}
\end{equation}
As a single photonic branch is considered, no $r,s$ indices are needed. A quantitative comparison of the two cases will be made in the final section Sec.\ref{sec:quant}.

\section{Emission rate from finite-size systems}
\label{sec:realistic}

In order to obtain a quantitative prediction for an actual experimental set-up, one
has to go beyond the idealized infinite geometry system considered so far, and study the more realistic case where dynamical Casimir light is generated in a spatially finite system and then revealed by a detector located in the external vacuum.

The first step is to rewrite the parametric Hamiltonian \eq{H_param} for a bulk system in a local form. Expressing the polariton operators $\ph_{r,\kk}$ in terms of the real-space ones
\begin{equation}
\ph_{r,\kk}=\frac{1}{\sqrt{\mc V}}\,\int\! d\xx\,e^{-i\kk\xx}\,\psih_r(\xx),
\end{equation}
the parametric Hamiltonian can be rewritten as:
\begin{multline}
\delta H=\frac{\hbar}{2}\,\sum_{rs}\int\!d\xx\,d\xx'\,
\psihd_r(\xx)\,\psihd_s(\xx')\,{\tilde V}_{rs}(\xx-\xx')\,e^{-i \omega t} \\ +\textrm{h.c.}.
\eqname{par_r}
\end{multline}
Here the kernel
\begin{equation}
{\tilde V}_{rs}(\xx)=\frac{1}{(2\pi)^3}\int\!d\kk\,e^{i\kk\xx}\,V_{rs,k}
\end{equation}
does not depend on the integration volume ${\mc V}$ and, under reasonable smoothness assumptions for $V_{rs,k}$, is a quite localized function around $\xx=0$.

\begin{figure}[htbp]
\begin{center}
\includegraphics[width=0.9\columnwidth,clip]{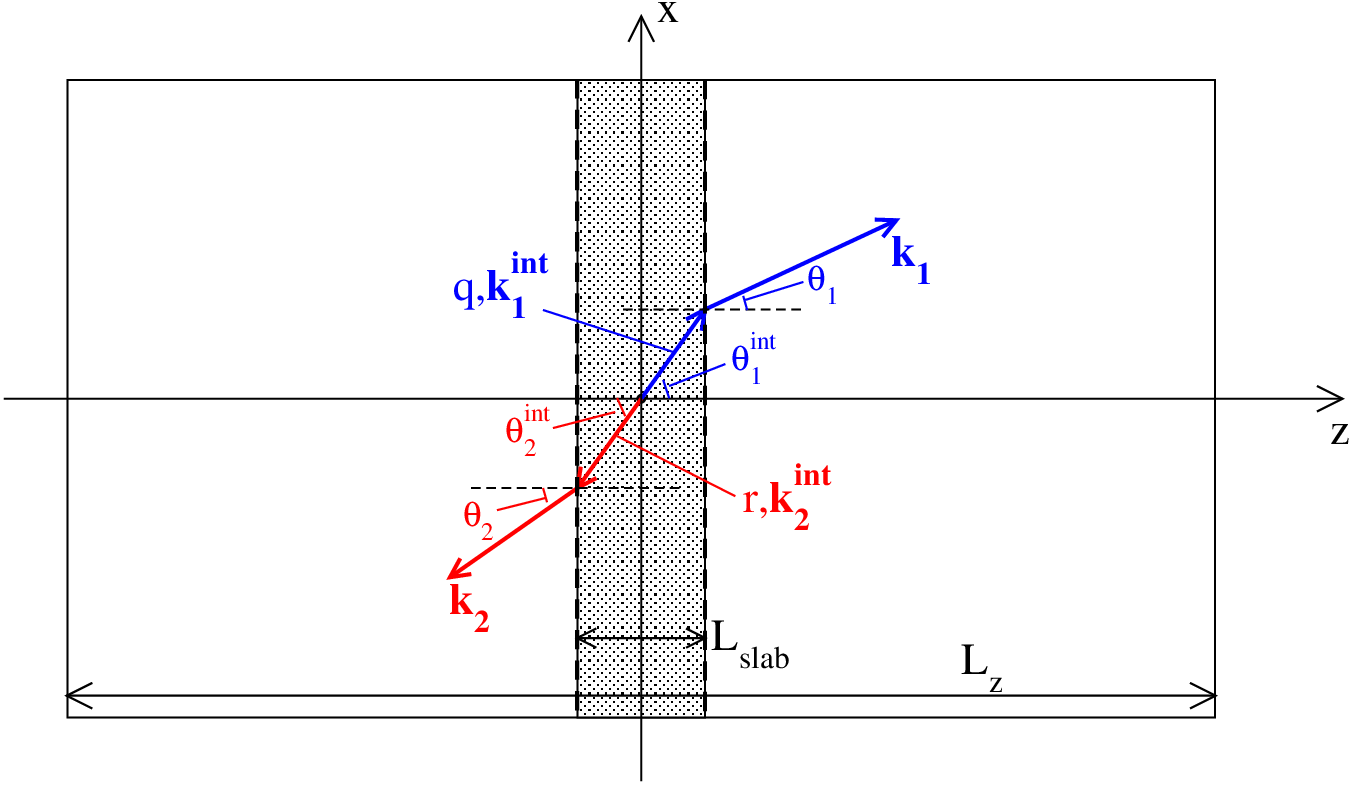}
\caption{Sketch of the slab geometry under consideration.}
\label{fig:slab}
\end{center}
\end{figure}

The advantage of the real-space Hamiltonian \eq{par_r} is that it is not limited to bulk systems, but can be also applied to finite geometries. For the sake of simplicity, we shall limit ourselves to the plane-parallel slab geometry of fig.\ref{fig:slab}.
The size $L_z$ of the integration box in the $z$ direction orthogonal to the slab plane is assumed to be much longer than the slab thickness $L_{slab}$. In the transverse $x,y$ directions, the slab is assumed to fill the whole integration box of size $L_{x,y}$.

In order to apply the Fermi golden rule, one has to identify the final states of the process.
In the present case, they consist of pairs of radiative photons emitted in the empty space surrounding the atomic system. These are created inside the system as polaritons, and then transform into free-space photons when traversing the interface to the external vacuum.
For the sake of simplicity, we shall assume that interface reflections are negligible at the frequencies of interest, so that the internal polariton is adiabatically transformed into the photon state at exactly the same energy.
As light is collected far outside the slab, we do not have to consider here the case of light being guided inside the slab by total internal reflection.

As usual in scattering problems, the final eigenstates are labelled by the wave vector $\kk$ outside the atomic system and their frequency is fixed by the free-space dispersion $\omega=ck$.
Outside the Mott insulator, polaritons reduce in fact to photons.
Their wave function is the plane wave:
\begin{equation}
\psi^{out}_{\kk}(\xx)=\frac{1}{\sqrt{{\mc V}}}\,e^{i\kk\xx},
\end{equation}
Inside the system, the same eigenstate corresponds to a polariton plane wave of the form
\footnote{In real space, the polariton wave function $\psi^{in}_{\kk}(\xx,r)$ depends in fact on the spatial coordinate $\xx$ and on the branch index $r$. This latter can be interpreted as a sort of internal degree of freedom like spin.}
\begin{equation}
\psi^{in}_{\kk}(\xx,r)=\frac{A}{\sqrt{{\mc V}}}\,\delta_{r,\bar{r}}\,e^{i\kk^{int}\xx}.
\end{equation}
The wavevector $\kk^{int}$ and the branch index $\bar{r}$ are fixed by energy conservation in the refraction process  at the interface separating the atomic system and the surrounding vacuum: the translational symmetry of the system along the $x,y$ in-plane directions guarantees that the $x,y$ components of the wave vector are conserved $k_{x,y}=k_{x,y}^{int}$, while the band index $\bar{r}$ and the $z$ component $k_z^{int}$ are fixed by the energy conservation $\omega_{\bar{r},k^{int}}=c k$
\footnote{As the bands have no spectral overlap, at most one propagating polariton mode exists for a given frequency $\omega$: this guarantees that no additional boundary conditions are required here~\cite{ABC}.
Rather, the energy conservation condition can be satisfied only outside the (very small) gaps that open in the polariton dispersion between the $LP$ and the $MP$, and between the $MP$ and the $UP$. For frequencies inside these gaps, the radiative eigenstates are confined in the external vacuum and can not penetrate the atomic system: the dynamical Casimir emission is therefore strongly suppressed. \\
On the other hand, light escape from the atomic system is possible only for polariton modes which lie within the so-called radiative cone $k_x^2+k_y^2\leq \omega^2/c^2$: polaritons which are created outside this cone remain in fact trapped in the slab and propagate through it as in a waveguide.}.

The amplitude $A$ also depends on the external wavevector $\kk$ and is fixed by the particle flux conservation condition. As interface reflections are assumed to be negligible, this reads:
\begin{equation}
c\,\cos(\theta)=|A|^2\,{\rm v}^{gr}_{\bar{r},k^{int}}\,\cos(\theta^{int}),
\end{equation}
where $\theta$ ($\theta^{int}$) are the angles between the $z$ axis and the propagation direction outside (inside) the slab: note the dramatic increase of the polariton amplitude inside the slab for ${\rm v}^{gr}\ll c$, a well-known effect in the theory of slow light propagation~\cite{fleischhauer}.
Provided the slab is much thinner than the integration box $L_z\gg L_{slab}$, it is important to remind that the normalization and the density of states do not depend on the slab size, and only involve the total volume ${\mc V}=L_x L_y L_z$.

Inserting the explicit form of the polariton wave function
$\psi_{\kk}(\xx,r)$ into the parametric Hamiltonian \eq{par_r} and
using the Fermi golden rule, one immediately gets to an expression for
the number of photons emitted into the external vacuum per unit time $dt$, unit surface $d\Sigma=dx\,dy$, and unit phase-space volume of transverse momentum $d^2k_\perp=dk_x\,dk_y$:
\begin{multline}
\frac{dN}{dt\,d\Sigma\,d^2k_\perp}= \\
=\frac{1}{(2\pi)^2}\;|V_{\bar{r}\bar{s},\bar{k}^{int}}|^2\,
\frac{L_{slab}}{\cos\theta^{int}\,({\rm v}^{gr}_{\bar{r},\bar{k}^{int}}+{\rm v}^{gr}_{\bar{s},\bar{k}^{int}})}.
\eqname{emission_slab}
\end{multline}
For each value of the modulation frequency $\omega$, the internal wave vector $\bar{k}^{int}$ and the branch indices $\bar{r},\bar{s}$ are selected by the resonance condition \eq{resonant} with $\kk$ replaced by $\kk^{int}$. 
Inside the system the wave vectors of the emitted polariton pair differ in fact from perfect anti-parallelism by a negligible amount $\Delta k_z\propto 1/L_{slab}$, still their frequencies $\omega_{(\bar{r},\bar{s}),\bar{k}^{int}}$ and their group velocities ${\rm v}^{gr}_{(\bar{r},\bar{s}),\bar{k}^{int}}$ can be significantly different as soon as distinct branches are considered $\bar{r}\neq \bar{s}$.

It is interesting to note that the result \eq{emission_slab} is in perfect agreement with the emission rate predicted in \eq{rate} for a bulk system of volume ${\mc V}_{slab}=L_xL_yL_{slab}$: in the absence of losses, all the polaritons created in the finite slab are in fact emitted from the system as radiation.

By means of a change of variables, the emission rate \eq{emission_slab} can be rewritten in its final form as an emission rate per unit surface and unit solid angle $d\Omega$:
\begin{multline}
\frac{dN_\pm}{dt\,d\Sigma\,d\Omega}=\frac{1}{(2\pi)^2}\;|V_{\bar{r}\bar{s},\bar{k}^{int}}|^2\,\\ 
\times \left(\frac{\omega_\pm}{c}\right)^2\,
\frac{\cos\theta_\pm}{\cos\theta^{int}}\,\frac{L_{slab}}{{\rm v}^{gr}_{\bar{r},\bar{k}^{int}}+{\rm v}^{gr}_{\bar{s},\bar{k}^{int}}},
\eqname{emission_slab2}
\end{multline}
where the $\pm$ index respectively refer to the photon which is emitted from the slab in the positive ($+$) or negative ($-$) $\hat{z}$ direction at angles respectively $\theta_\pm$ with the normal. This means e.g. that $\omega_+=\omega_{\bar{r},\bar{k}^{int}}$ and $\omega_-=\omega_{\bar{s},\bar{k}^{int}}$. The difference in the angular emission density in the $\pm$ directions is due to refraction effects of the $\bar{r},\bar{s}$ polaritons at the system-vacuum interfaces.

\subsection{Resonant enhancement in a cavity}
\label{sec:cavity}

A possible way to further enhance the emission intensity is to surround the slab with a pair of mirrors of good reflectivity $R \lesssim 1$: if the modulation frequency $\omega$ is on resonance with a pair of cavity modes, the dynamical Casimir emission results increased by a factor proportional to the finesse of the cavity~\cite{lambrecht}.

For simplicity, let us consider a plane-parallel mirror geometry, so that the cavity modes are labeled by the in-plane component $\kk_\perp$ of the wave vector, the branch index $\bar{r}$, as well as by a positive integer number $M$ defining the mode order along $z$. Assuming for simplicity that the mirrors are metallic and that no vacuum space is left between the slab and the mirrors, the polaritonic cavity modes are defined by the round-trip quantization condition: 
\begin{equation}
k^{int}_z\,L_{slab}=\pi\,M,
\end{equation}
and their frequency is equal to the bulk polariton dispersion $\omega_{\bar{r},k^{int}}$ at the relevant wave vector $k^{int}$.
Because of the non-trivial shape of the dispersion law $\omega_{\bar{r},k^{int}}$ as a function of $k^{int}$, the cavity modes are not equally spaced in frequency. 
Their polaritonic wave function has a simple sinusoidal form (the mirrors are at $z=0$, $L_{slab}$):
\begin{equation}
\psi_{\bar{r},M,\kk_\perp}(\xx,r)=\sqrt{\frac{2}{L_{slab}L_xL_y}}\,
\sin\left(\frac{\pi M\,z}{L_{slab}}\right)\,\delta_{r,\bar{r}}\,e^{i\kk_\perp\cdot\xx}
\end{equation}
and their radiative decay rate into externally propagating photons is equal to
\begin{equation}
\gamma=\frac{1-R}{L_{slab}}\,\cos(\theta^{int})\,{\rm v}^{gr}_{\bar{r},k^{int}}.
\eqname{gammas}
\end{equation}
As expected, the slower the group velocity ${\rm v}^{gr}$, the smaller the decay rate $\gamma$.

Fermi golden rule can again be used to estimate the emission rate in the cavity geometry. As final states, pairs of cavity photons have to be considered, with a finite line width equal to $\gamma$~\cite{CCT4}.
For a thick enough cavity, the mode order at the frequency of interest is $M\gg 1$ and overlap factor strongly privileges polariton emission into pairs of cavity modes of the same order $M$; the efficiency of all other processes is suppressed by their spatial phase mismatch. 
The emission rate into a pair of such modes is then easily obtained:
\begin{multline}
\frac{dN}{dt\,d\Sigma\,d^2k_\perp}=\frac{1}{\pi^2}\,|V_{\bar{r}\bar{s},\bar{k}^{int}}|^2\,
\frac{\gamma_T}{(\omega-\omega_T)^2+\frac{\gamma_T^2}{4}} \\
\stackrel{\omega=\omega_T}{\longrightarrow} \frac{1}{\pi^2\,\gamma_T}\,|V_{\bar{r}\bar{s},\bar{k}^{int}}|^2,
\eqname{emission_cavity}
\end{multline}
where $\omega_T$ and $\gamma_T$ are here the sum of respectively the frequencies and the linewidths of the pair of modes under examination.

By comparing equation \eq{emission_cavity} with the result \eq{emission_slab} in the absence of the enclosing cavity, and using \eq{gammas}, it is immediate to see that the emission rate for an excitation exactly on resonance with a pair of modes ($\omega=\omega_T$) is enhanced by a factor $4/(1-R)$. This enhancement effect is quite general and holds for a variety of optical processes~\cite{cavity_enh}; for well reflecting mirrors $R\lesssim 1$, it can be quite dramatic.

\section{Quantitative discussion and experimental considerations}
\label{sec:quant}

In the previous sections, we have obtained simple analytical expressions relating the emission intensity in the different geometries to the $V_{rs,k}$ parameter \eq{V_qrk} which carries informations on the microscopic optical properties of the atomic medium. In the present section, we shall conclude the study by providing quantitative estimation of the emission intensity for realistic systems, and discussing the most relevant issues that are likely to arise in the design of an actual experiment.

\begin{figure}[htbp]
\begin{center}
\includegraphics[width=\columnwidth,clip]{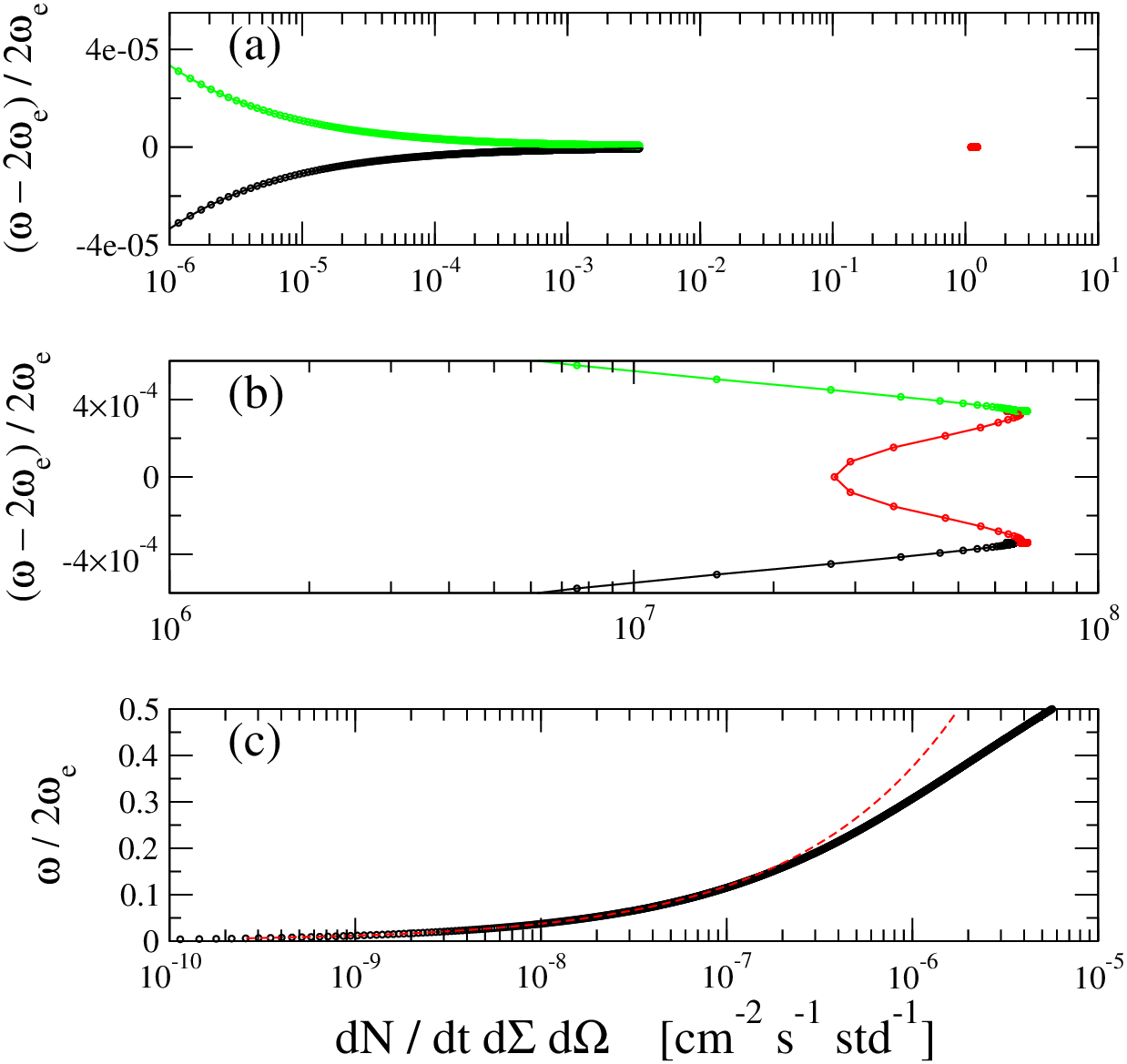}
\caption{
Spectrum of the dynamical Casimir emission rate per unit surface and unit solid angle around the normal direction ($\theta_\pm=0$) as a function of the modulation frequency $\omega$ for the resonant case $\omega_e=\tilde{\omega}_m$, a slab thickness $L_{slab}=10\,\mu\textrm{m}$ in the absence of enclosing cavity, and a relative modulation amplitude $\delta\Omega_C/\Omega_C=0.05$.
Same system parameters as in Fig.\ref{fig:polar_disp}.
(a) weak dressing $\Omega_C/{\bar C}= 1.86\cdot 10^{-4}\ll 1$; (b,c) strong dressing $\Omega_C/\bar{C}=2$.
The circles in (a,b,c) are the result of the exact calculation \eq{emission_slab2}. The solid lines in (a,b) are the analytical approximation \eq{analytic1}. The red dashed line in (c) is the prediction of the dispersionless, time-varying dielectric model based on \eq{V_k_diel}. Panels (a,b) correspond to the band dispersion and absorption spectra shown in Fig.\ref{fig:polar_disp}(a-d). 
}
\label{fig:spectrum}
\end{center}
\end{figure}

\begin{figure}[htbp]
\begin{center}
\includegraphics[width=0.9\columnwidth,clip]{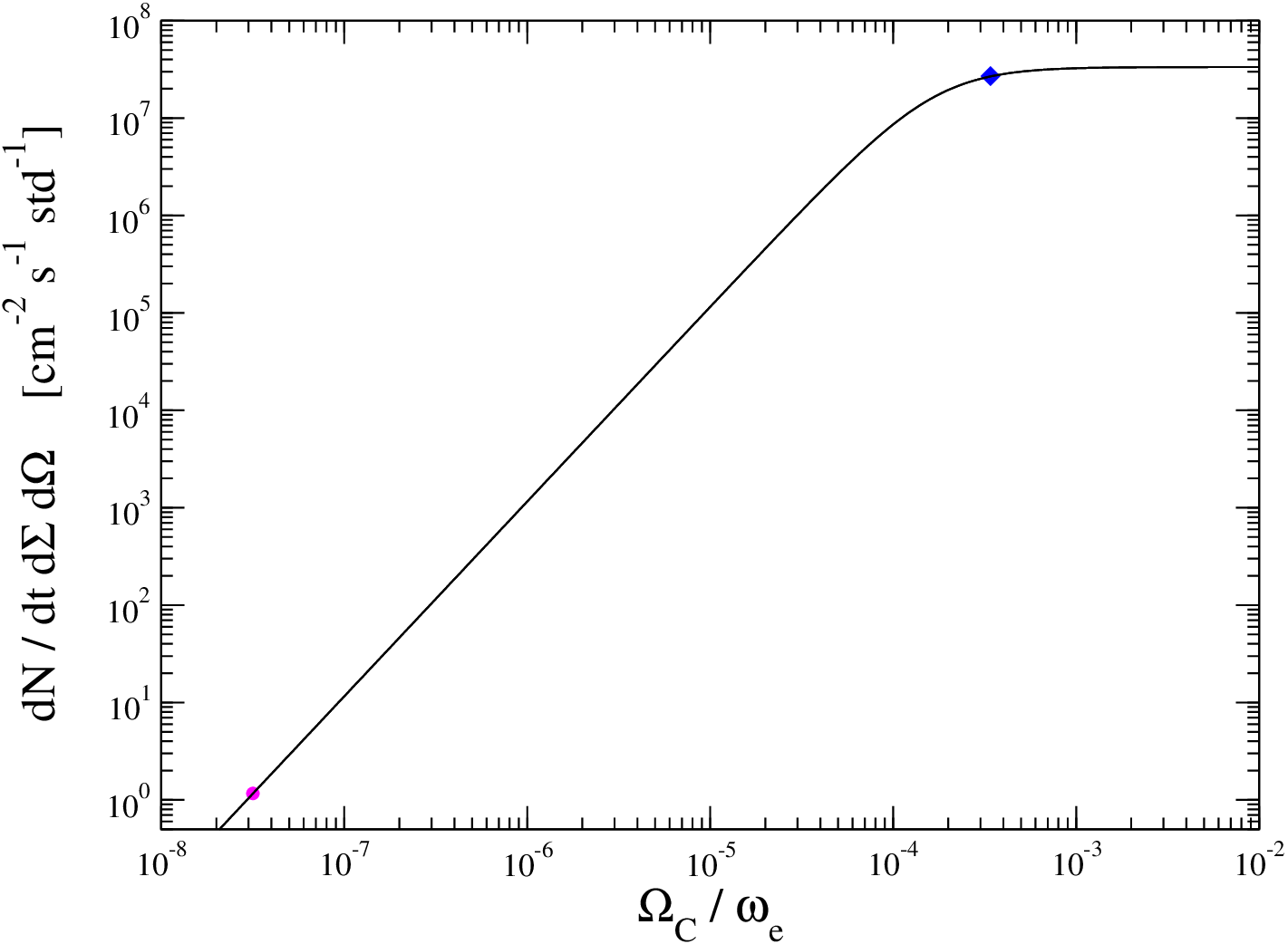}
\caption{Prediction \eq{emission_slab2} for the dynamical Casimir emission rate per unit surface and unit solid angle around the normal direction ($\theta_{\pm}=0$) as a function of the coupling amplitude $\Omega_C$ for a fixed value of the relative modulation amplitude $\delta\Omega_C/\Omega_C=0.05$ at the fully resonant point $\omega/2=\omega_e=\tilde{\omega}_m$. 
Same system parameters as in Fig.\ref{fig:spectrum}: the magenta circle and the blue diamond correspond to respectively the weak (a) and strong (b) dressing regimes.
}
\label{fig:intensity}
\end{center}
\end{figure}

An example of emission intensity spectrum is shown in Fig.\ref{fig:spectrum} for a $L_{slab}=10\,\mu$m slab of Rb atomic Mott insulator in the absence of cavity as described by \eq{emission_slab2}.
The coupling amplitude $\Omega_C$ is assumed to have the periodic time-dependence \eq{Omega_C} at a frequency $\omega$, while its frequency $\omega_C$ is kept fixed: the modulation of $\Omega_C$ then consists of a pair of coherent sidebands at $\pm 2(\omega_e-\omega_g)$ around the carrier frequency $\omega_C=|\omega_e-\omega_m|$.
As expected, the central $MP$ polariton branch appears as the most favorable region thanks to the combination of reduced group velocity and significant resonant mixing of light and matter excitations.
In the resonant region [panels (a,b)] the result of the complete calculation is compared with the analytical approximation \eq{analytic1}: the agreement is excellent.
In panel (c), the same calculation is performed in the low frequency region and is compared to Law's result \eq{V_k_diel} for a dispersionless medium~\cite{Law}: inserting in \eq{V_k_diel} the refractive index variation that follows from time dependence of $\Omega_C$ and using the general formula \eq{emission_slab2} for the emission rate, one obtains the dashed curve in Fig.\ref{fig:intensity}c.  
As expected, at very low frequencies the agreement is excellent, but dispersion effects start playing a significant role already for $\omega\approx 0.2\,\omega_{e}$.

All panels of Fig.\ref{fig:spectrum} have been calculated for the same value of the modulation amplitude $\delta \Omega_C/\Omega_C$: in spite of the higher value of ${\rm v}^{gr}$, a larger value of $\Omega_C$ is favourable in view of maximizing the emission intensity as it allows for a stronger modulation of the optical properties.
A specific plot of the resonance emission rate as a function of the coupling amplitude $\Omega_C$ is shown in Fig.\ref{fig:intensity}: as predicted in Sec.\ref{sec:matrix_element}, the Casimir emission first grows as $\Omega_C^2$ for $\Omega_C\ll \bar{C}$, then saturates to a finite value for $\Omega_C\gg\bar{C}$.

Remarkably, the emission intensity at resonance can reach quite substantial values already in the absence of a cavity.
In the weak dressing case ($\Omega_C/2\pi=12\,\textrm{MHz}$), the rate of emitted photons from a $1\,\textrm{cm}^2$ system in the unit solid angle around the normal is of the order of $1$ photon per second [see Fig.\ref{fig:spectrum}(a)].
Then it quadratically (see Fig.\ref{fig:intensity}) increases for growing $\Omega_C$, to eventually saturate around a value larger than $10^7$  photon per second for a huge dressing $\Omega_C/2\pi\simeq 100\,\textrm{GHz}$ [see Fig.\ref{fig:spectrum}(b)].
For alkali atoms such as Rb, a dressing amplitude $\Omega_C/2\pi$ in the $10\,\textrm{MHz}$ range corresponds to intensities of the dressing beam in the $\textrm{mW}/\textrm{cm}^2$ range~\cite{steck}.

A crucial difficulty of most dynamical Casimir experiments consists of varying the optical properties of the system at a high enough speed. In our specific setup, this amounts to modulating the dressing beam at a frequency which is resonant with the creation of a pair of MP.

Although this is hardly done with the almost symmetric $\Lambda$ schemes currently used in slow light experiments with alkali atom samples~\cite{Arimondo,SlowLightExp,stop}, still strongly asymmetric $\Lambda$ schemes as shown in Fig.\ref{fig:levels}(c) can be used.
The ground state of the atom is now the $m$ state, while the $g$ state is a long-lived, high-energy metastable state.
The dressing beam then acts on the $m\rightarrow e$ transition at a frequency $|\omega_e-\omega_m|$ higher than twice the frequency $|\omega_e-\omega_g|$ of the $g\rightarrow e$ transition on which the dynamical Casimir radiation is to be emitted: no principle difficulties then appear to prevent one from modulating the dressing amplitude $\Omega_C$ at the required frequency $\omega\simeq 2\,|\omega_e-\omega_g|$
\footnote{A dressing amplitude $\Omega_C(t)$ modulated at the optical frequency $\omega$ can be obtained e.g. by nonlinear wave mixing on a suitable nonlinear crystal of a carrier at $\omega_C$ with another beam at $\omega$.}.

Suitable strongly asymmetric $\Lambda$ configurations can be found e.g. in alkali-earth atoms~\cite{alkaliearthbook} whose laser cooling and trapping techniques have experienced remarkable advances in the last few years~\cite{alkaliearth}.
A specific choice in view of our dynamical Casimir application can be $^{88}$Sr atoms, whose $4d\,^1D_2$ metastable state appears to have the required properties to be used as the $g$ state of a strongly asymmetric $\Lambda$ scheme.
It is in fact connected to the excited $5p\,^1P_1$ state by an optically active infrared transition at $\lambda=6.5\,\mu$m which, for a lattice spacing of 300nm, gives a value $\bar{C}/\omega_e\simeq 5\cdot 10^{-5}$ not far from the Rb one.
The $5p\,^1P_1$ state can be dressed by driving the atom on the $\lambda=461\,$nm transition from the absolute ground state of the atom $5s^2\,^1S_0\rightarrow 5p\,^1P_1$.
Before performing the dynamical Casimir experiment, atoms have to be optically pumped in the $4d\,^1D_2$ metastable state e.g. by means of a $\pi$ Raman pulse; the lifetime of the state being of $0.33$ms, there is enough time left to carry out the dynamical Casimir experiment before atoms decay to the ground state via the $5p\,^3P_J$ state: even for a group velocity as low as $10\,$m/s, the transit time across a $10\,\mu$m thick cloud is in fact much shorter, of the order of $1\,\mu$s. 
Note also that this spontaneous decay channel involves photons at $1.8\,\mu$m and $689\,$nm, in completely different spectral regions from the dynamical Casimir ones at $6.5\,\mu$m, which can therefore be spectrally isolated with no difficulty.

Another, may-be simpler solution is to stick to alkali atoms such as Rb, but use a ladder scheme [see Fig.\ref{fig:levels}(b)] where the $m$ state is a electronically highly excited state instead of a $\Lambda$ one.
Even though the unavoidable radiative contribution to $\gamma_m$ prevents the EIT effect from completely killing the absorption, still DCE light can escape the slab without dramatic losses. This, at least for the large values of $\Omega_C$ of present interest [Fig.\ref{fig:MP_summ}(d)].

As the EIT effect is not crucial for the observation of the dynamical Casimir emission, one may think of doing experiment with atomic samples where spontaneous emission from both the $e$ and the $m$ states is still effective, e.g. Bose-Einstein condensates or even thermal gases.
Additional difficulties could however arise in such systems from the inhomogeneous broadening of $\tilde{\omega}_m$ due to the spatial variations of the trapping and interaction potentials and from the reduced value of the light-matter coupling coefficient $\bar{C}$ due to the reduced atomic density.
While this latter effect can be quite dramatic in non-degenerate clouds, the inhomogeneous broadening of $\tilde{\omega}_m$ should be in both cases overcome by choosing a sufficiently strong coupling amplitude $\Omega_C$.

\section{Conclusions}
\label{sec:conclu}

In this paper we have performed a complete systematic analysis of the optical properties of a gas of coherently dressed three-level atoms trapped in an optical lattice in a Mott insulator state.
The extreme degree of coherence of this system allows for propagation of light at ultra-slow group velocities for long times and distances.
The optical properties of the medium can be controlled in real time by varying the amplitude and the frequency of the dressing field. For sufficiently fast modulation rates, the zero-point fluctuations of the polariton vacuum state are converted into observable radiation by dynamical Casimir effect.

We have developed a general theory to quantitatively characterize the dynamical Casimir emission in terms of a simple parametric Hamiltonian and we have identified the most favourable case of a resonant dressing frequency whose amplitude is periodically modulated in time.  Experimentally realistic geometries such as plane-parallel slabs and planar cavities are analysed in detail. Remarkably, a sizeable radiation intensity is predicted for state-of-the-art systems and no spurious emission from blackbody radiation or incoherent luminescence is expected to mask the dynamical Casimir signal.

\begin{acknowledgments}

We are grateful to G. C. La Rocca, M. Artoni, R. Sturani, G. Ferrari, C. Tozzo, F. Dalfovo for stimulating discussions.
IC acknowledges hospitality at the  Institut Henri Poincare-Centre Emile Borel and financial support from the french CNRS.

\end{acknowledgments}

\end{document}